\documentclass[12pt,preprint]{aastex}
\usepackage{xcolor}
\usepackage{soul}
\usepackage{epsfig}
\definecolor{armygreen}{rgb}{0.29, 0.33, 0.13}
\definecolor{darkorchid}{rgb}{0.6, 0.2, 0.8}
\definecolor{skyblue}{rgb}{0.53, 0.81, 0.92}

\begin{document}

\title{Probing the Broad Line Region and the Accretion Disk in the Lensed Quasars HE0435-1223, WFI2033-4723, and HE2149-2745 using Gravitational Microlensing}

\author{V. Motta}
\affil{Instituto de F\'{\i}sica y Astronom\'{\i}a, Universidad de Valpara\'{\i}so, Avda. Gran Breta\~na 1111, Playa Ancha, Valpara\'{\i}so 2360102, Chile} 

\author{E. Mediavilla}
\affil{Instituto de Astrof\'{\i}sica de Canarias, Universidad de La Laguna, Avda. V\'{\i}a L\'actea s/n, La Laguna, Tenerife 38200, Spain}
\affil{Departamento de Astrof\'{\i}sica, Universidad de La Laguna, La Laguna, Tenerife 38205, Spain}

\author{K. Rojas}
\affil{Instituto de F\'{\i}sica y Astronom\'{\i}a, Universidad de Valpara\'{\i}so, Avda. Gran Breta\~na 1111, Playa Ancha, Valpara\'{\i}so 2360102, Chile} 

\author{E. E. Falco}
\affil{Harvard-Smithsonian Center for Astrophysics, 60 Garden St, Cambridge, MA 02138, USA}

\author{J. Jim\'enez-Vicente}
\affil{Departamento de F\'{\i}sica Te\'orica y del Cosmos, Universidad de Granada, Campus de Fuentenueva, 18071 Granada, Spain\\ Instituto Carlos I de F\'{\i}sica Te\'orica y Computacional, Universidad de Granada, E-18071 Granada, Spain} 

\and

\author{J.A. Mu\~noz}
\affil{Departamento de Astronom\'{\i}a y Astrof\'{\i}sica, 
Universidad de Valencia, E-46100 Burjassot, Valencia, Spain}
\affil{Observatorio Astron\'omico, Universidad de Valencia, E-46980 Paterna, Valencia, Spain} 

\begin{abstract}

We use single-epoch spectroscopy of three gravitationally lensed quasars, HE0435-1223, WFI2033-4723, and HE2149-2745, to study their inner structure (BLR and continuum source).  We detect microlensing-induced magnification in the wings of the broad emission lines of two of the systems (HE0435-1223 and WFI2033-4723). In the case of WFI2033-4723, microlensing affects two ``bumps'' in the spectra  which are almost symmetrically arranged on the blue (coincident with an \ion{Al}{3}  emission line) and red wings of \ion{C}{3]}. These match the typical double-peaked profile   that follows from disk kinematics. The presence of microlensing in  the wings of the emission lines indicates    the existence of two different regions in the BLR: a    relatively small one with kinematics possibly related to an accretion   disk, and another one that is substantially more extended and insensitive to   microlensing. There is good agreement between   the estimated size of the region affected by microlensing in the   emission lines, $r_s=10^{+15}_{-7} \sqrt{M/M_{\odot}}$~light-days   (red wing of \ion{C}{4} in HE0435-1223) and $r_s=11^{+28}_{-7}   \sqrt{M/M_{\odot}}$ light-days (\ion{C}{3]} bumps in WFI2033-4723) with     the sizes inferred from the continuum emission, $r_s=13^{+5}_{-4} \sqrt{M/M_{\odot}}$~light-days (HE0435-1223)     and $r_s=10^{+3}_{-2} \sqrt{M/M_{\odot}}$~light-days     (WFI2033-4723). For HE2149-2745 we measure an accretion disk     size $r_s=8^{+11}_{-5} \sqrt{M/M_{\odot}}$~light-days. The     estimates of $p$, the exponent of the size vs. wavelength     ($r_s\propto\lambda^p$), are $1.2\pm0.6$, $0.8\pm0.2$, and     $0.4\pm0.3$ for HE0435-1223, WFI2033-4723, and HE2149-2745,      respectively.  In conclusion, the continuum microlensing amplitude     in the three quasars and chromaticity in WFI2033-4723 and     HE2149-2745 are below expectations for the thin disk     model. The disks are larger and their temperature     gradients are flatter than predicted by this model.

\end{abstract}


\section{Introduction} 

Gravitationally lensed quasars are very well suited to study the inner structure of AGN \citep{pooley07,anguita08,bate08,eigenbrod2008,poindexter08,chartas09,floyd09,dai10,morgan10,blackburne11,mosquera11a,munoz11,chen12,hainline12,morgan12,hainline13,blackburne14,jimenez14,jimenez15a,macleod15,mediavilla15a,mediavilla15b,munoz16}. The magnification of lensed quasar images depends on the mass distribution of the lens galaxy and on geometrical considerations (distances and alignment between quasar, galaxy and observer). Therefore, the mean (or macromodel) flux ratios between images would be very useful observables to study the mass density profile in lens galaxies, assuming that all the lens galaxy mass is smoothly distributed. However, compact objects such as stars in the lens galaxies induce strong spatial gradients in the gravitational potential that give rise to anomalies in the flux ratios compared to the predictions of macromodels. This effect \cite[so-called microlensing; see][]{chang79,wambsganss06} complicates the macro-modeling of lens systems but in exchange it has a very useful property: it is sensitive to the source size (with smaller source regions showing larger magnifications). Thus, we can study the size of the emitting region by measuring flux ratios  to deduce the effects of microlensing.

Taking advantage of this property, the inner structure of AGNs can be
studied by searching for the effects of microlensing on different regions
of single-epoch spectra of the images of a lensed quasar
\citep{mediavilla11,motta12,guerras13,rojas14,sluse15}. The cores of
the emission lines, which likely arise from extended regions, is
insensitive to microlensing, so their flux ratios are 
a baseline against which we can measure the effects of
microlensing on other regions of smaller size, such as the wings of the
emission lines \citep{richards04,motta12,sluse12,guerras13,braibant14}
and the continuum generated by the accretion disk.  According to the
thin disk model \citep{shakura73}, the size of the continuum
varies with wavelength with a $r_s\propto \lambda^{4/3}$ law, and
hence the microlensing magnification will show some wavelength
dependence (so-called microlensing chromaticity).

The objective of this paper is to use single-epoch spectra of three
gravitationally lensed quasars (HE0435-1223, WFI2033-4723, and
HE2149-2745) to discuss the presence of microlensing in the emission
lines (which would yield information on the structure of the BLR) and in the
continuum (to estimate sizes and temperature gradients in the
accretion disk). The paper is organized as follows. In section 2 we
present the data. Section 3 is devoted to a description of our  
method of analysis.  We discuss our results in \S4 and offer 
concluding remarks in section \S5.

\section{Observations and data reduction}

HE0435-12223 was observed on January 12th 2008 with the Blue Channel
spectrograph on the MMT. Spectroscopic information for WFI2033-4723 and 
HE2149-2745 was gathered 
during April-May 2008 with the FORS2 spectrograph at the Very Large
Telescope (VLT)\footnote{based on observations made with ESO
  telescopes at Paranal Observatory under program 381.A-0508(A),
  P.I. V. Motta}.  
Table \ref{obs} summarizes the main observational characteristics of the data. 
For HE0435-1223, we also
used archival data\footnote{based on data obtained from the ESO
  Science Archive Facility, program 074.A-0563(B).} obtained with the
FORS1 spectrograph on the VLT. For the three systems we also analyzed
the deconvolved spectra from \cite{sluse12} provided by the VizieR
\citep{vizier} catalogue\footnote{based on data obtained with the Vizier
  catalogue access tool, CDS, Strasbourg, France.}.  A detailed
description of these observations and the spectrum analysis can be
found in \cite{eigenbrod2007} and \cite{sluse12}.

\begin{deluxetable}{lccccccccr}
\tabletypesize{\scriptsize}
\tablewidth{0pt}
\tablecaption{Log of observations \label{obs}}
\tablehead{
  \colhead{Objects} &
  \colhead{Pair\tablenotemark{a}} &
  \colhead{$\Delta$\tablenotemark{b} (\arcsec)} &
  \colhead{Instrument} &
  \colhead{Grating} &
  \colhead{Date} &
  \colhead{Airmass} &
  \colhead{P.A.\tablenotemark{c}} &
  \colhead{Seeing\tablenotemark{d}} &
  \colhead{Exposure\tablenotemark{e}}
}
\startdata
HE0435-1223   & BD & 1.5 & MMT/Blue-Channel & 300   & 2008/01/12 & 1.40 & -13.92 & 0.87 & 1800 \\
              & BD\tablenotemark{f} &  & VLT/FORS1   & 300   & 2004/10/11 &  &  & & 1400  \\ 
              & BD\tablenotemark{f} &  & VLT/FORS1   & 300   & 2004/10/12 &  &  & & 1400  \\
              & BD\tablenotemark{f} &  & VLT/FORS1   & 300   & 2004/11/11 &  &  & & 1400  \\
WFI2033-4723  & BC\tablenotemark{h} & 2.1 & VLT/FORS2 & 300   & 2008/04/14 & 1.23 & -79.05 & 0.7 & $3\times720$\\
              & BC\tablenotemark{g} &  & VLT/FORS2  & 300 & 2005/05/13 & 1.16 &   & 0.54 & $5\times1400$\\
HE2149-2745   & AB\tablenotemark{h} & 1.7 & VLT/FORS2 & 300   & 2008/05/07 & 1.38 & -28.63 & 0.8 & $3\times300$\\
              & AB\tablenotemark{g} &  & VLT/FORS2 & 300   & 2006/08/04 & 1.48 & -32.0 & 0.62 & $6\times1400$\\
\enddata
\tablenotetext{a}{Pair or image observed}
\tablenotetext{b}{Separation between images in arcsec}
\tablenotetext{c}{Position angle in degrees E of N}
\tablenotetext{d}{Seeing in arcseconds}
\tablenotetext{e}{Seconds of time}
\tablenotetext{f}{Data obtained from the ESO Science Archive Facility from program 074.A-0563(B), P.I. G. Meyland \citep{eigenbrod2007}}
\tablenotetext{g}{Deconvolved spectra from \cite{sluse12} (VizieR Archive) based on \cite{eigenbrod2007} data.}
\tablenotetext{h}{Observations made with ESO telescopes at Paranal Observatory under program 381.A-0508(A), P.I. V. Motta \citep{motta12}}
\end{deluxetable}

Data reductions were carried out with IRAF\footnote{IRAF is distributed
  by the National Optical Astronomy Observatory, which is operated by
  the Association of Universities for Research in Astronomy, Inc.,
  under cooperative agreement with the National Science Foundation}
tasks. 
The procedure consisted of subtraction, flat fielding, and wavelength calibration. 
As we are interested only in flux ratios between the quasar images, flux-calibration is not needed.
The cosmic-rays were removed using at least three exposures. 
The 1-D spectra
extraction is obtained by simultaneously fitting two Gaussian
functions to the components for each wavelength. The deconvolved
spectra obtained from the VizieR archive are already fully reduced.
The systematic errors affecting our measurements were discussed
elsewhere \citep{motta12}.

\section{Methods}

The procedure we use to separate microlensing and extinction consist of  
measuring the displacement between the continuum and the core
of the emission line flux ratios \citep[e.g.,
  see][]{mediavilla09,mediavilla11,motta12,guerras13,rojas14}.  
Thus, the baseline for no-microlensing is established by using the line core fluxes. 
 The continuum is retrieved by fitting the regions on either side of each
emission line ($\lambda_i$ to $\lambda_f$ wavelength range) and its flux is the integral below such function ($F_c$).  
 For instance, when $y_c = a \lambda + b$ is used as fitting function, the integrated flux is obtained as $F_c=(a/2)(\lambda_f-\lambda_i)^2+b(\lambda_f-\lambda_i)$. 
Once the continuum is subtracted, the core flux is obtained by
integrating the emission line profiles using DIPSO \citep{howarth04} in
STARLINK\footnote{Support provided by the Starlink Project which is
  run by CCLRC on behalf of PPARC.}.  
The integration is performed in a narrow interval (from 20 to
100~\AA \ depending on the line profile shape) centered on the peak of line. 
Narrower integration windows are chosen in those cases in which absorption lines are present (e.g. 20~\AA \ for \ion{C}{4} in HE2149-2745 \ion{C}{4}). 
The continuum fitting error ($\Delta a$, $\Delta b$) provided by DIPSO (at 1$\sigma$ level)\footnote{errors in the linear approximation are calculated from the error matrix} is used as an estimation of the core flux error. Specifically, the errors for each coefficient ($\Delta a$, $\Delta b$) are used to estimate the error in the continuum as  $\Delta F_c =(\Delta a/2)(\lambda_f-\lambda_i)^2+\Delta b (\lambda_f - \lambda_i)$. As the core flux measurement relies on the continuum fitting, its error is estimated as the error in the continuum \citep[see][]{motta12}. 

We also compare our results with magnitude differences obtained from
the literature (Table \ref{lit}), measured in the near-infrared. Longer wavelengths are
expected to be less affected by microlensing because they are produced
in a larger emitting region (however, as \cite{fadely11} have stated,
for sources with $z_s<2.8$ the $L$ broad-band could be contaminated by
thermal emission from the inner dusty torus)

\begin{deluxetable}{lcccccr}
\tabletypesize{\scriptsize}
\tablecaption{Summary of Known Quasar Image Fluxes \label{lit}}
\tablewidth{0pt} 
\tablehead{ 
  \colhead{Lens Name} &
  \colhead{z$_L$\tablenotemark{a}} &
  \colhead{z$_S$\tablenotemark{b}} &
  \colhead{Filter\tablenotemark{c}} &
  \colhead{$1/\lambda$  \tablenotemark{d} ($\mu$m$^{-1}$)} &
  \colhead{$\Delta m$ (mag) \tablenotemark{e}} &
  \colhead{Source \tablenotemark{f}}
}
\startdata
HE0435-1223 BD & 0.46 & 1.689  & L'           & 0.26 &  $-0.22 \pm 0.09$  & 5 \\
               &      &        & K            & 0.45 &  $-0.26 \pm 0.02$  & 5 \\
               &      &        & Ks           & 0.45 &  $-0.26 \pm 0.02$  & 9 \\
               &      &        & H            & 0.61 &  $-0.20 \pm 0.03$  & 9 \\
               &      &        & F160W        & 0.65 &  $-0.26 \pm 0.02$  & 1 \\
               &      &        & J            & 0.80 &  $-0.19 \pm 0.04$  & 9 \\
               &      &        & z'           & 1.00 &  $-0.20 \pm 0.02$  & 9 \\
               &      &        & F814W        & 1.23 &  $-0.23 \pm 0.04$  & 1 \\
               &      &        & I            & 1.27 &  $-0.22 \pm 0.04$  & 4\\
               &      &        & I            & 1.27 &  $-0.19 \pm 0.05$  & 4\\
               &      &        & i            & 1.30 &  $-0.15 \pm 0.03$  & 2 \\
               &      &        & i'           & 1.30 &  $-0.22 \pm 0.03$  & 9 \\
               &      &        & Iac29        & 1.43 &  $-0.24 \pm 0.06$  & 4 \\
               &      &        & H$\alpha$    & 1.52 &  $-0.22 \pm 0.02$  & 4 \\
               &      &        & H$\alpha$    & 1.52 &  $-0.23 \pm 0.02$  & 4 \\
               &      &        & r            & 1.60 &  $-0.13 \pm 0.03$  & 2 \\
               &      &        & r'           & 1.60 &  $-0.13 \pm 0.03$  & 9 \\
               &      &        & Iac28        & 1.65 &  $-0.20 \pm 0.07$  & 4 \\
               &      &        & r            & 1.60 &  $-0.02 \pm 0.02$  & 3 \\
               &      &        & F555W        & 1.80 &  $-0.13 \pm 0.06$  & 1 \\
               &      &        & V            & 1.83 &  $-0.02 \pm 0.02$  & 3 \\
               &      &        & Str-y        & 1.83 &  $-0.16 \pm 0.01$  & 4 \\
               &      &        & Str-y        & 1.83 &  $-0.21 \pm 0.07$  & 4 \\
               &      &        & g            & 2.08 &  $-0.23 \pm 0.03$  & 2 \\
               &      &        & g'           & 2.08 &  $-0.08 \pm 0.03$  & 9 \\
               &      &        & g            & 2.08 &  $-0.04 \pm 0.02$  & 3 \\
               &      &        & Str-b        & 2.14 &  $-0.13 \pm 0.06$  & 4 \\
               &      &        & Str-v        & 2.43 &  $-0.15 \pm 0.03$  & 4 \\
               &      &        & Str-v        & 2.43 &  $-0.24 \pm 0.02$  & 4 \\
               &      &        & u'           & 2.84 &  $0.11 \pm 0.02$   & 9 \\
               &      &        & Str-u        & 2.85 &  $-0.06 \pm 0.13$  & 4 \\
               &      &        & 0.5-0.8KeV   & 3.23 &  $-0.03 \pm 0.21$  & 9 \\
               &      &        & 0.4-0.8KeV   & 3.23 &  $-0.03 \pm 0.2$  & 12\\
WFI2033-4723 BC & 0.66 & 1.66  & Ks           & 0.75 &  $0.10 \pm 0.03$  & 9 \\
               &      &        & H            & 1.00 &  $0.13 \pm 0.07$  & 9 \\
               &      &        & F160W        & 1.07 &  $0.05 \pm 0.03$  & 1 \\
               &      &        & J            & 1.33 &  $0.15 \pm 0.02$  & 9 \\
               &      &        & z'           & 1.82 &  $0.28 \pm 0.03$  & 9 \\
               &      &        & F814W        & 2.07 &  $0.19 \pm 0.14$  & 1 \\
               &      &        & i'           & 2.16 &  $0.23 \pm 0.02$  & 9 \\
               &      &        & i'           & 2.16 &  $0.09 \pm 0.01$  & 10 \\
               &      &        & r            & 2.57 &  $0.17 \pm 0.05$  & 11 \\
               &      &        & r'           & 2.65 &  $0.38 \pm 0.02$  & 9 \\
               &      &        & r'           & 2.65 &  $0.12 \pm 0.01$  & 10 \\
               &      &        & F555W        & 3.05 &  $0.29 \pm 0.04$  & 1 \\
               &      &        & g'           & 3.46 &  $0.48 \pm 0.02$  & 9 \\
               &      &        & g'           & 3.46 &  $0.15 \pm 0.01$  & 10 \\
               &      &        & u'           & 4.71 &  $0.70 \pm 0.03$  & 9 \\
               &      &        & u'           & 4.71 &  $0.31 \pm 0.01$  & 10 \\
               &      &        & 0.5-0.8KeV   & 3.23 &  $0.49 \pm 0.17$  & 10 \\
HE2149-2745 AB & 0.60 & 2.032  & L'           & 0.26 &  $0.24 \pm 0.01$   & 5 \\
               &      &        & K            & 0.45 &  $0.28 \pm 0.06$   & 5 \\
               &      &        & F160W        & 0.65 &  $1.56 \pm 0.04$   & 1 \\
               &      &        & F814W        & 1.23 &  $1.56 \pm 0.02$   & 1 \\
               &      &        & i            & 1.30 &  $1.505 \pm 0.003$ & 8 \\
               &      &        & R            & 1.55 &  $1.57 \pm 0.04$   & 6\\
               &      &        & R            & 1.55 &  $1.59 \pm 0.02$   & 7\\
               &      &        & V            & 1.83 &  $1.57 \pm 0.06$   & 7 \\
               &      &        & V            & 1.83 &  $1.635 \pm 0.001$ & 8 \\
               &      &        & V            & 1.83 &  $1.64 \pm 0.04$   & 8 \\
               &      &        & F555W        & 1.80 &  $1.70 \pm 0.02$   & 1 \\
               &      &        & B            & 2.28 &  $1.62 \pm 0.06$   & 6\\
               &      &        & B            & 2.28 &  $1.57 \pm 0.03$   & 7\\
\enddata
\tablenotetext{a}{~Lens galaxy redshift}
\tablenotetext{b}{~Lensed quasar redshift}
\tablenotetext{c}{~Filter or, when available, we give the line emission flux 
between parenthesis}
\tablenotetext{d}{~Inverse of the central wavelength (rest frame). 
}
\tablenotetext{e}{~Magnitude difference between pair of images.} 
\tablenotetext{f}{~REFERENCES: 
(1) CASTLES; 
(2) \cite{wisotzki02}; 
(3) \cite{wisotzki03}; 
(4) \cite{mosquera11b}; 
(5) \cite{fadely11}; 
(6) \cite{wisotzki96}; 
(7) \cite{lopez98}; 
(8) \cite{burud02}; 
(9) \cite{blackburne11}; 
(10) \cite{morgan04}; 
(11) \cite{vuissoz08}; 
(12) \cite{chen12};
}
\end{deluxetable}

In those cases where our spectra are contaminated by the lens
galaxy continuum (e.g. HE2149-2745), we compare our line core flux
ratio with the uncontaminated broad-band measurements
(e.g. CASTLES\footnote{http://www.cfa.harvard.edu/glensdata/}).

Chromatic microlensing detection allows us to study the structure of
the accretion disk in the lensed quasar by estimating its size and
temperature profile. 
The accretion disk is modeled as a Gaussian,
$I\propto \exp(-R^2/2r_s^2)$, with size 
$r_s\propto \lambda^p$. 
\cite{mortonson05} showed that microlensing primarily constrains the half-light
radius of the disk $R_{1/2}$, but the
precise details of the disk model are unimportant, because  
the Gaussian scale length $r_s$ we use here is related
to the half-light radius by $R_{1/2}=1.18 r_s$ and this can be used to compare our
results to other choices for the disk profile.
The probability of reproducing the measured microlensing magnifications 
is estimated by randomly placing a Gaussian source on microlensing magnification 
maps. 
The Inverse Polygon Mapping method \citep{mediavilla06,mediavilla11} was used to compute maps of $2000
\times 2000$ pixels of 0.5~light days for each image.
Microlensing magnification statistics is degenerate with the single-mass case except in the case of a mass function with a high dynamical range, when a similar contribution to the mass density from microlenses at the opposite ends exists (bimodality). Previous works \citep{witt93,wyithe00,congdon07} support that microlensing magnification does not depend on the details of the mass function, but rather on the  mean mass of the  microlenses through the scaling factor $\sqrt{<M>}$, at least for Salpeter's like laws (but see also \cite{mediavilla15a} for more general stellar mass functions). 
Thus, we use $1 \ M_{\odot}$ microlenses and the size scales
linearly with this mass, i.e. $\propto
\sqrt{<M>/M_{\odot}}$ .  The convergence ($\kappa$) and shear
($\gamma$) for each image are selected using macromodels from the
literature \citep{mediavilla09,sluse12}.  The fraction of mass in
compact objects is assumed to be $\alpha=0.1$, an acceptable value considering 
current estimates \cite[see
  e.g.][]{schechter02,mediavilla09,pooley09,jimenez15a,jimenez15b}.
Bayesian posterior probabilities 
\citep{mediavilla11} of $r_s$ and $p$ conditioned to the measured microlensing
magnifications are estimated using a uniform logarithmic (linear) prior on $r_s$ ($p$).

\section{Results} \label{results}

\subsection{HE0435-1223} \label{he0435}

Discovered by \cite{wisotzki02}, it consists of four images of a
lensed quasar at $z=1.689$ and a lens galaxy at $z_L=0.455$
\citep{morgan05,ofek2006}. CASTLES images obtained with HST also
revealed multiple partial arcs between the images.  We present spectra for
the $B$ and $D$ images taken with MMT and VLT telescopes. Due to the
symmetric distribution of the images, the time delay between images is
small \cite[$<10$ days;][]{kochanek06,courbin11}.

\begin{figure*}
\begin{center}
\includegraphics[width=5cm]{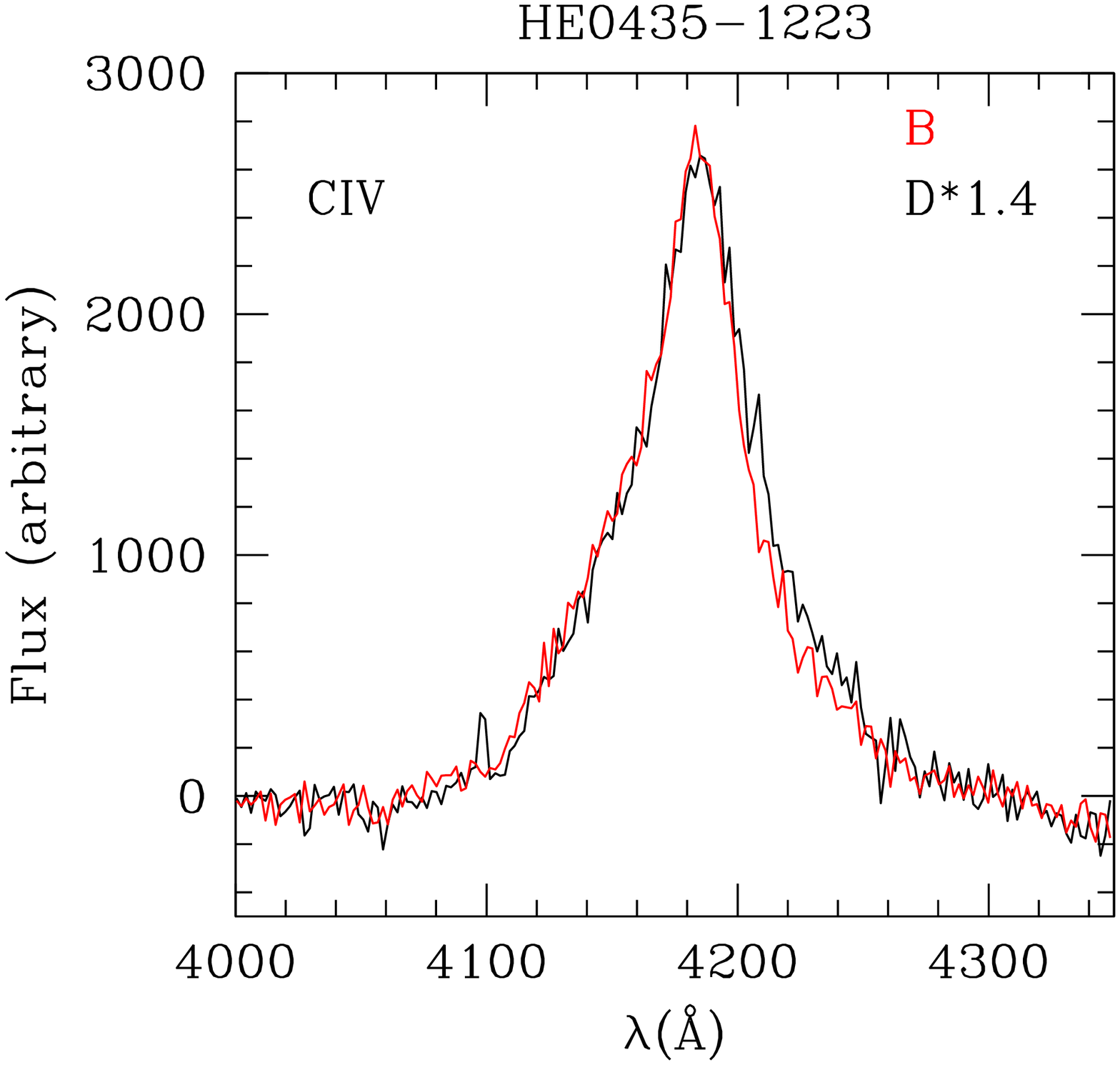}
\includegraphics[width=5cm]{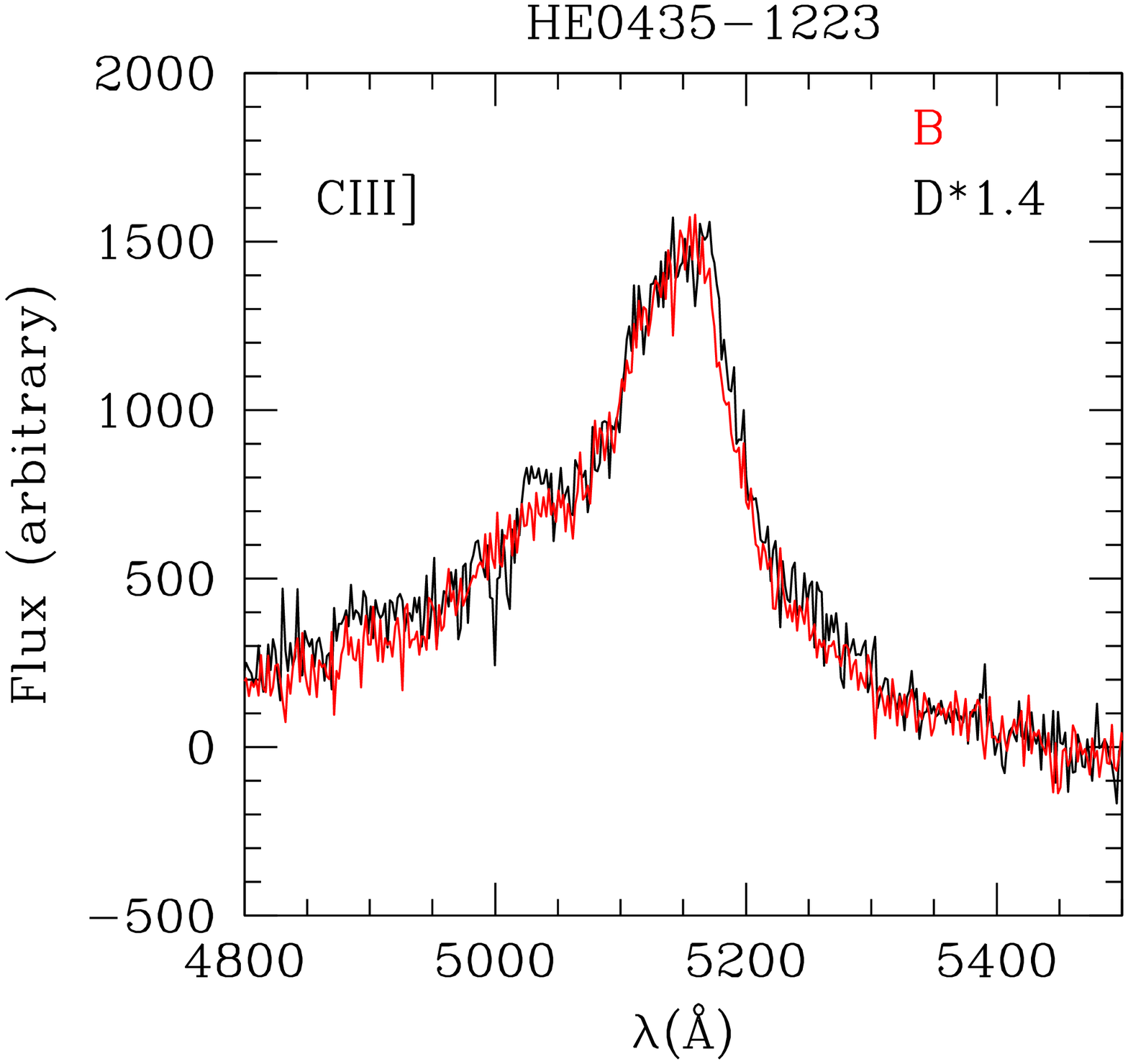}
\includegraphics[width=5cm]{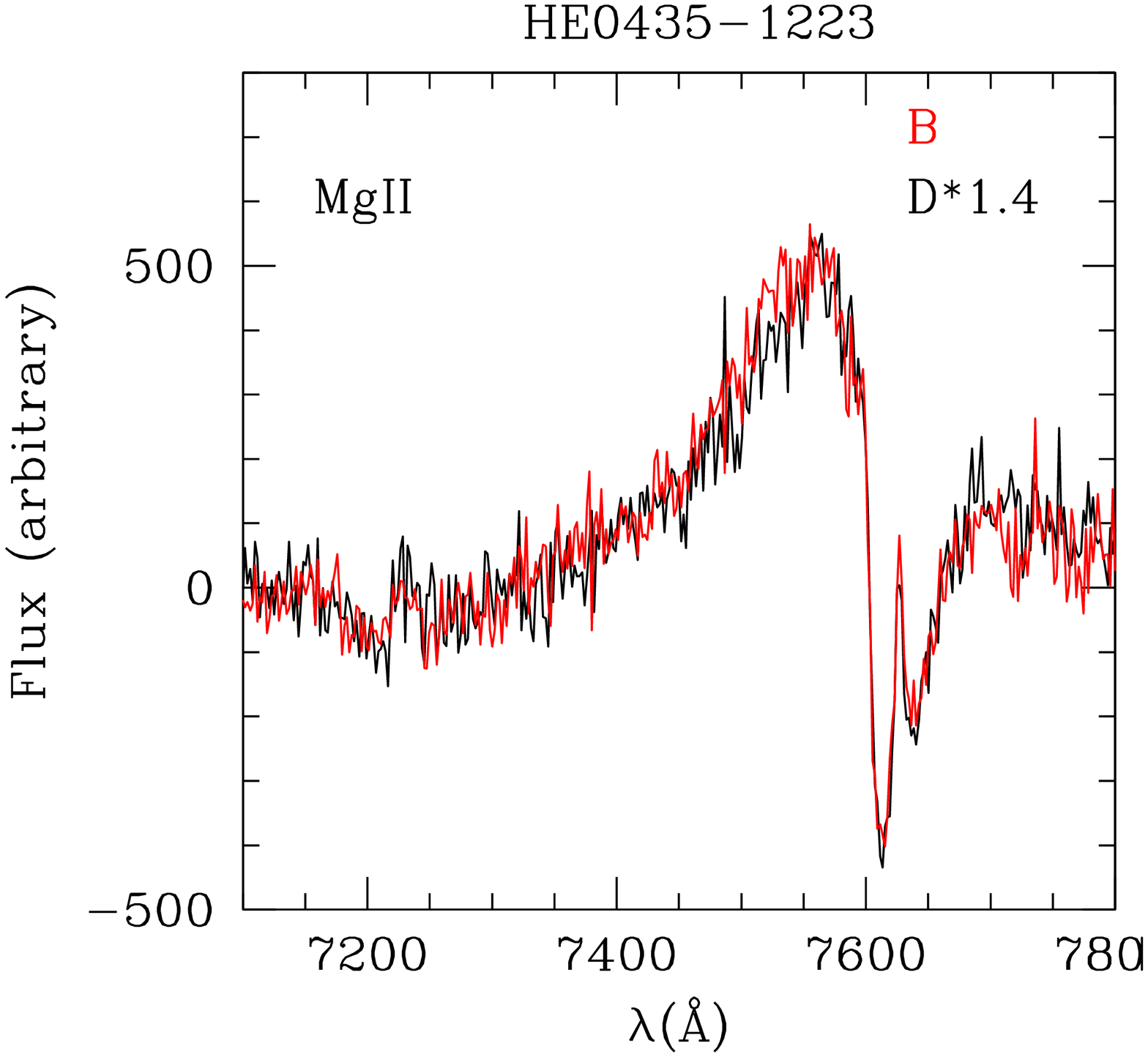}\\
\includegraphics[width=5cm]{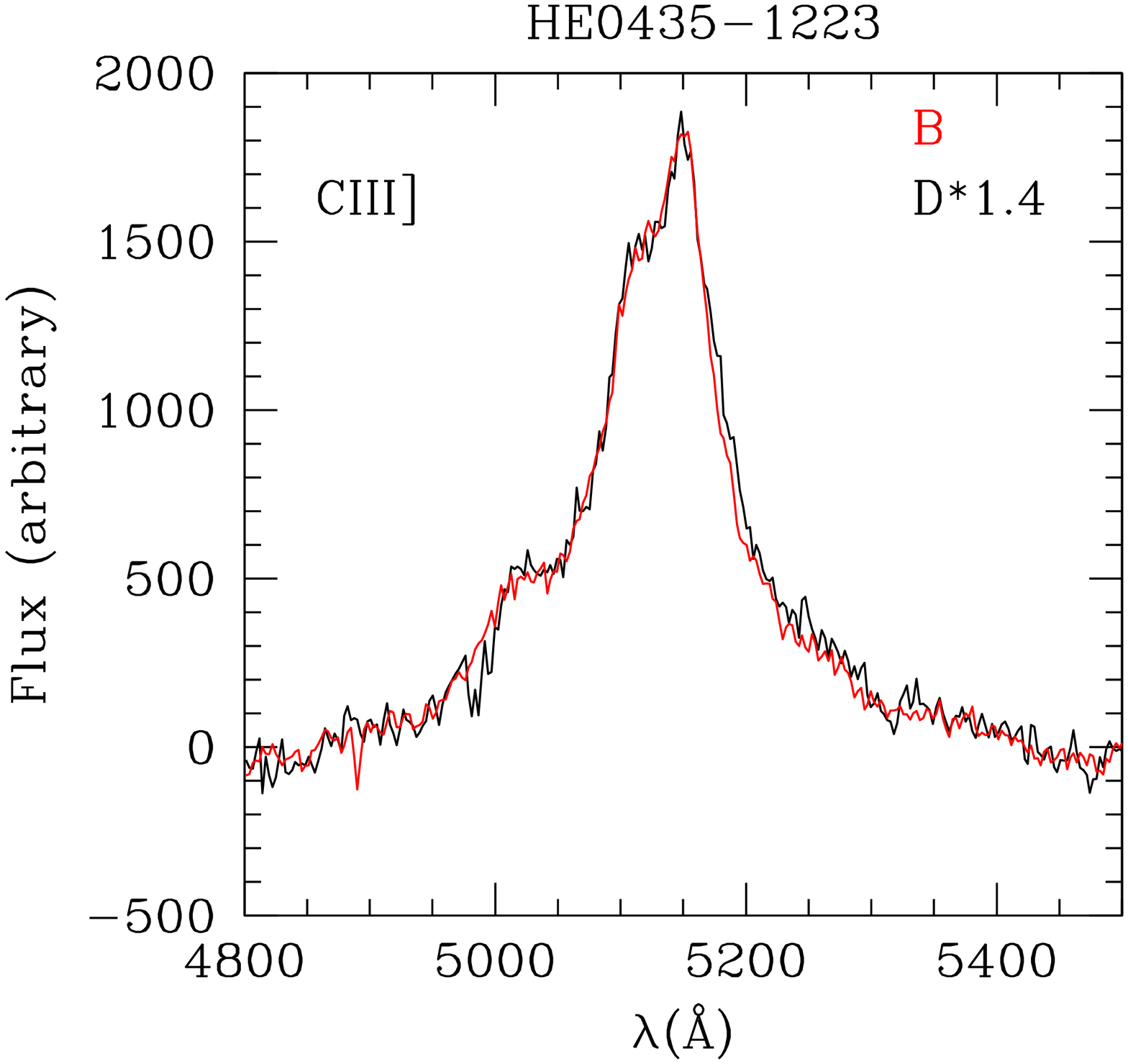}
\includegraphics[width=5cm]{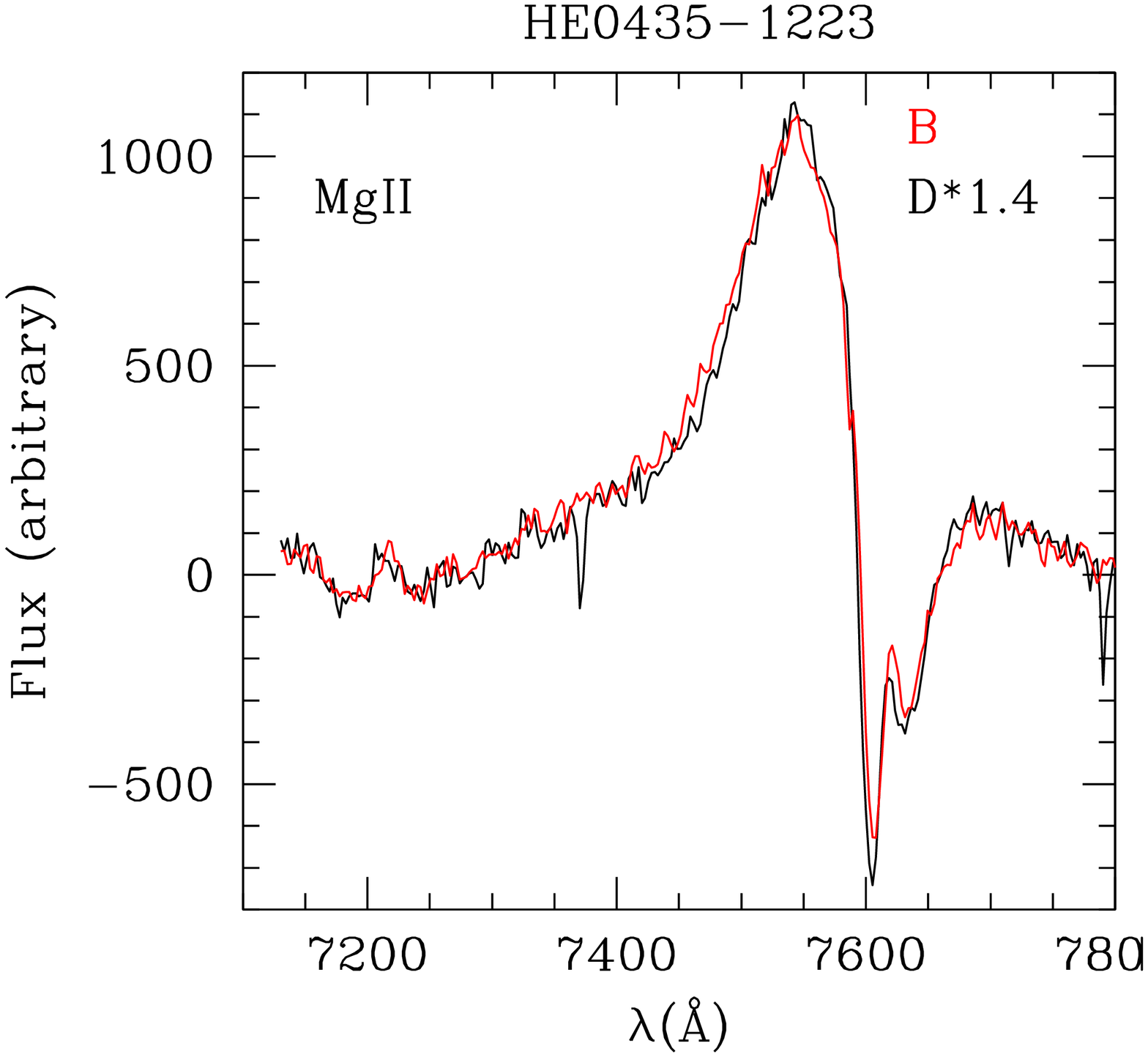}\\
\caption{\ion{C}{4}, \ion{C}{3]}, and \ion{Mg}{2} emission line profile for HE0435-1223B,D vs. observed $\lambda$ for 
MMT ({\em top}) and VLT ({\em bottom}) data.
The {\em red line} represents the continuum subtracted emission lines for image $B$. 
The {\em black line} represents the continuum subtracted emission line for image $D$ 
multiplied by a factor to match the peak of $B$. The factors are shown in each panel. 
\label{prof_he0435bd}}
\end{center}
\end{figure*}

In Figure \ref{prof_he0435bd} we present the continuum subtracted spectra in the regions
corresponding to the \ion{C}{4}, \ion{C}{3]} and \ion{Mg}{2} emission lines. The spectra
  are normalized to match the profiles in the
  region of the core of the line (see \S 3).  A very interesting
  result in the high ionization \ion{C}{4} line is the presence of a slight
  but systematic enhancement of the red wing ($\sim [4200,4250]$~\AA
  \, range) of component $D$ with respect to component $B$.  A similar
  enhancement of the red wing is hinted at in component $A$ (not plotted
  here) with respect to component $B$, but with less strength. 
  \ion{C}{4} profiles of images $B$ and $C$ (not plotted here)
  seem to match reasonably well.  In the case of the low ionization
  \ion{C}{3]} line, the red wing of image $D$ also seems to be brighter than
    that of image $B$, but by a small amount. Finally, the lower
    ionization line \ion{Mg}{2} shows no significant differences between
    image profiles.

Differences present at specific wavelengths in the line profiles and
the dependence of its strength on the degree of ionization
(decreasing in the \ion{C}{4}, \ion{C}{3]}, \ion{Mg}{2} sequence) can be explained as 
  microlensing acting selectively on parts of a broad line region with
  organized kinematics. $B$ and $C$ would be the images less affected
  by differential microlensing and $D$ the most affected.

From near IR observations of the H$\alpha$ emission line,
\cite{braibant14} also propose microlensing to explain the differences
between profiles. They found that the H$\alpha$ emission line profiles
of images $B$ and $C$ match well and that the effect of microlensing
seems to be more pronounced for image $D$. However, their Figure 1 shows an
excess in the image $B$ profile in the blue wing relative to $D$
(instead of the red wing excess of $D$ that we measure). This is likely
due to the different approach they followed to compare the
profiles. They normalize the continuum before superposing
them. Thus, their normalization factors include macro-magnification
and continuum microlensing. In this way, they are mixing these two
effects with line microlensing in the comparison of two emission line
profiles. Following our procedure, on the contrary, continuum
microlensing is removed by continuum subtraction, and
macro-magnification is corrected for by normalization to the core of
the emission line (assumed to be insensitive to microlensing).  In
fact, the re-normalization of the emission line profiles of
\cite{braibant14} to match the core of the lines would likely result
in an enhancement of the red wing of the $D$ emission line profile with
respect to that of $B$.  In the same manner, the blue wing enhancements
reported by \cite{braibant14} in the \ion{Mg}{2} emission line will also
likely disappear after normalization to the line cores.

Integrating the red wing excess in \ion{C}{4} corresponding to image $D$ and
$B$ in the $[4200,4250]$~\AA \, range, we obtain the microlensing
magnification associated with the region. The red wing magnification was
obtained as \cite[see][]{guerras13}: $\Delta m^{red\,
  wing}_{BD}=(m_B-m_D)_{[4200,4250]}-(m_B-m_D)_{core}=(-0.12\pm0.03)-(-0.37\pm0.02)=(0.25\pm0.04)$~mag.

\begin{figure*}
\begin{center}
\plotone{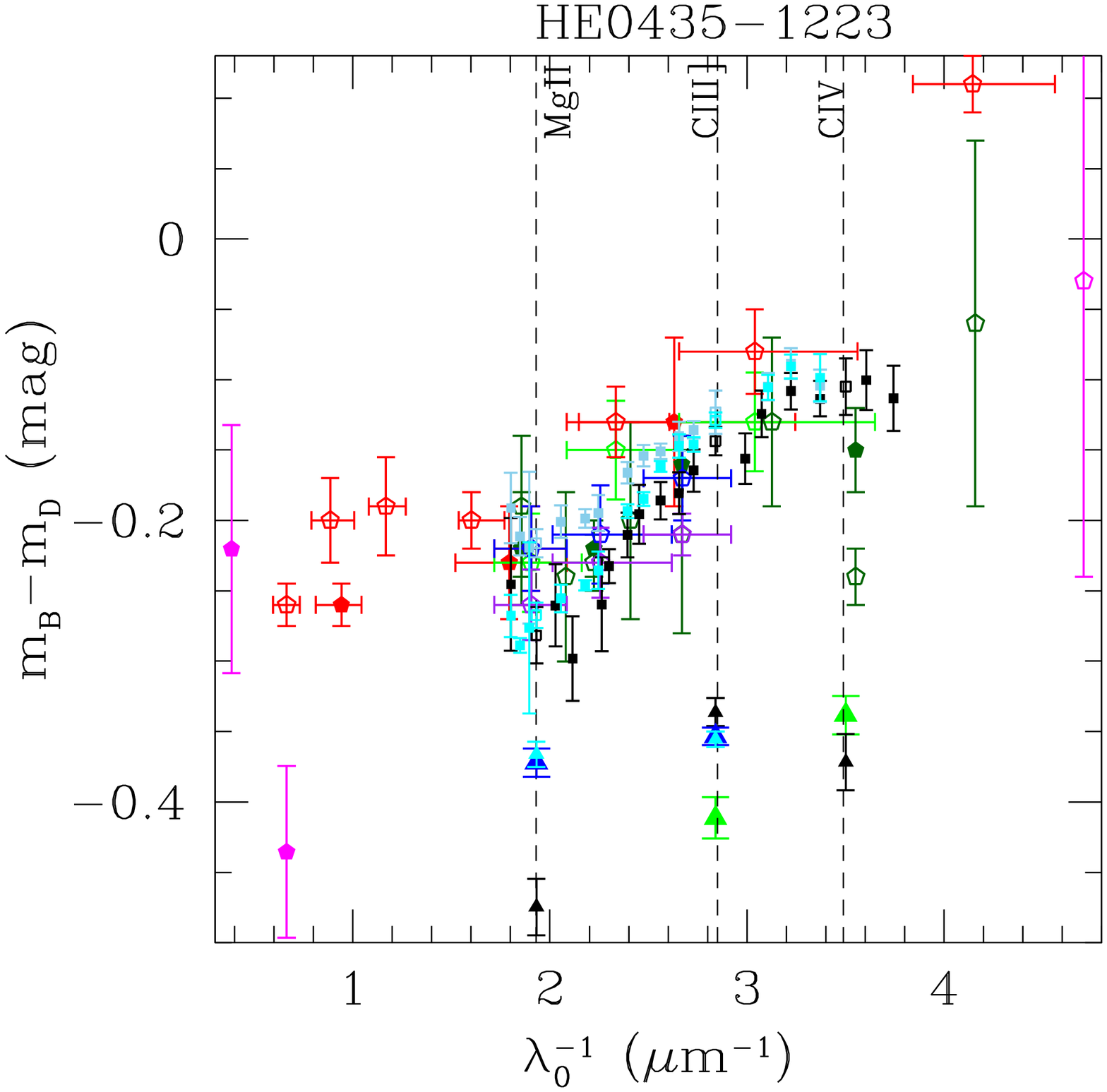}
\caption{\footnotesize{ Magnitude differences $m_B-m_D$ vs $\lambda_0^{-1}$ ($\lambda$ in the lens galaxy restframe) for  HE0435-1223. 
We use the standard units of $\mu$m$^{-1}$ for extinction studies, within the range of observed $\lambda$. {\em Pentagons} represent the integrated continuum obtained from (broad-band) CASTLES ({\em solid red}), \cite{wisotzki03} ({\em solid green}), \cite{wisotzki02} ({\em open green}), \cite{blackburne11} ({\em open red}), \cite{fadely11} ({\em solid magenta}),  \cite{mosquera11b} ({\em open} and {\em solid dark green}), \cite{ricci11} (epoch 2008 {\em open violet}, epoch 2009 {\em open cyan}). 
The {\em squares} represent the magnitude differences from the integrated continuum in our spectra ({\em solid black}) and from the integrated fitted continuum under the emission lines ({\em open black}) for MMT, VLT archive ({\em solid blue}, {\em open blue}, and VLT deconvolved data ({\em solid cyan}, {\em open cyan}) respectively. 
{\em Triangles} are the magnitude difference in the emission line cores ({\em solid black}, {\em solid blue}, {\em solid cyan}) for MMT, VLT archive and VLT deconvolved spectra respectively. 
For display convenience, average X-ray data \citep{blackburne11,chen12} ({\em open magenta pentagons}) shifted in wavelength to 4.7$\mu$m$^{-1}$ (i.e. from $\sim600\mu$m$^{-1}$ to $\sim4.7\mu$m$^{-1}$ in the rest frame). 
}
\label{diff_he0435a}}
\end{center}
\end{figure*}

In spite of the presence of microlensing in the red wings of some line
profiles, the ratio of the line cores (see Figure \ref{diff_he0435a}
and Table \ref{mag_he0435}) agrees within uncertainties. We obtain an
average value for the emission lines ratio of 
$\Delta m_L=<m_B-m_D>_L=-0.37\pm0.01$~mag. This confirms the expectation that
the cores are not very sensitive to microlensing and that extinction
is not significantly present in this lens system \cite[in agreement
  with][]{wisotzki03,morgan05}.

\begin{deluxetable}{llcrr}
\tablecolumns{5}
\tablewidth{0pt} 
\tablecaption{HE0435-1223 magnitude differences \label{mag_he0435}}
\tablehead{ 
  \colhead{Region} &
  \colhead{$\lambda_c$ (\AA)} &
  \colhead{Window\tablenotemark{c} (\AA)} &
  \colhead{$m_D-m_B$\tablenotemark{a} (mag)} &
  \colhead{$m_D-m_B$\tablenotemark{b} (mag)} 
}
\startdata
Continuum & 4170     & 4000-4350 & $-0.11\pm0.02$  & \nodata        \\
          & 5140     & 4540-5550 & $-0.14\pm0.01$  & $-0.12\pm0.02$ \\
          & 7560     & 7130-7880 & $-0.28\pm0.02$  & $-0.22\pm0.01$ \\
\hline             
Line  & \ion{C}{4}$\lambda$1549              & 4170-4195 & $-0.37\pm0.02$ & \nodata \\
      & \ion{C}{4} red wing                  & 4200-4250 & $-0.12\pm0.03$ & \nodata \\
      & \ion{C}{3]}$\lambda$1909            & 5100-5180 & $-0.34\pm0.01$ & $-0.36\pm0.01$ \\
      & \ion{Mg}{2}$\lambda$2800             & 7480-7580 & $-0.47\pm0.02$ & $-0.37\pm0.01$ \\
\enddata
\tablenotetext{a}{MMT data} 
\tablenotetext{b}{VLT archive data}
\tablenotetext{c}{Integration window.}
\end{deluxetable}

Continuum observations also indicate that, in contrast with 
$B$, $A$ and $D$ are affected by microlensing. \cite{courbin11} conclude that
the $B$ image is the least affected by stellar microlensing while $A$
is affected by strong microlensing variations. \cite{wisotzki03}
found evidence of chromatic microlensing in $D$ component (0.07~mag)
between Dec. 2001 and Sep. 2002. Using lightcurves in the R filter,
\cite{kochanek06} also observed microlensing in $D$ (relative to $A$)
of $\sim 0.1$~mag~yr$^{-1}$. \cite{mosquera11b} obtained
lightcurves with narrow-band filters in Oct. 2007, observing a
chromaticity between the bluest (Str-b) and the reddest (I-band)
filters of $\Delta m_{I-b}=0.20\pm0.09$~mag affecting $A$
component. \cite{ricci11} observed the system in the i, V, R bands in
two epochs concluding that image $A$ is probably affected by
microlensing.

In Figure \ref{diff_he0435a} we show the continuum ratio
values available from the literature and those obtained from the
spectra used in the present work. In the wavelength region between
\ion{Mg}{2} and \ion{C}{4} all these data consistently depict a linear trend with a
variable gap with respect to the no microlensing baseline defined by
the emission line ratios of about 0.1 and 0.3~mag respectively at the
red (\ion{Mg}{2}) and blue (\ion{C}{4}) ends. From \ion{C}{4} towards the UV, the broad
band data available seem to fit in this linear trend albeit with a
large scatter. From \ion{Mg}{2} towards the IR, the chromaticity disappears
and, with the exception of the outlying K band data from
\cite{fadely11}, the broad band data seem to concentrate around a
constant offset of $\sim$0.18~mag with respect to the baseline defined
by the emission line ratios.

\begin{figure*}
\begin{center}
\plotone{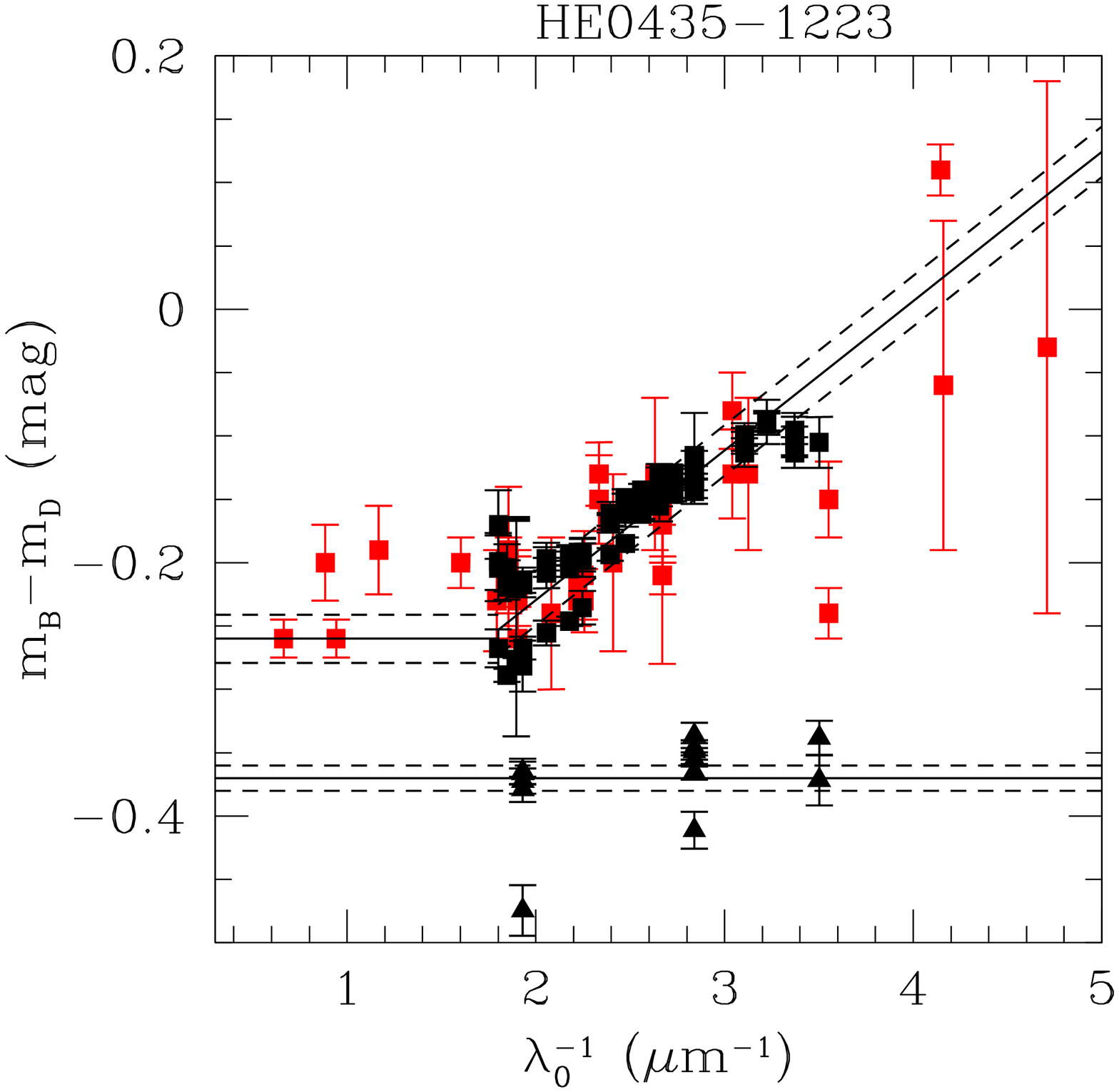}
\caption{Models fitted to the data shown in Figure \ref{diff_he0435a}.  
{\em Squares} and {\em triangles} represent continuum and line core data respectively.  
{\em Continuum lines} represent the function fitted to the continua and the average of the emission lines. 
{\em Dashed lines} are the standard deviations for the continuum fit and the standard 
error of the mean for the emission lines. 
\label{diff_he0435b}}
\end{center}
\end{figure*}

\begin{figure*}
\begin{center}
\plotone{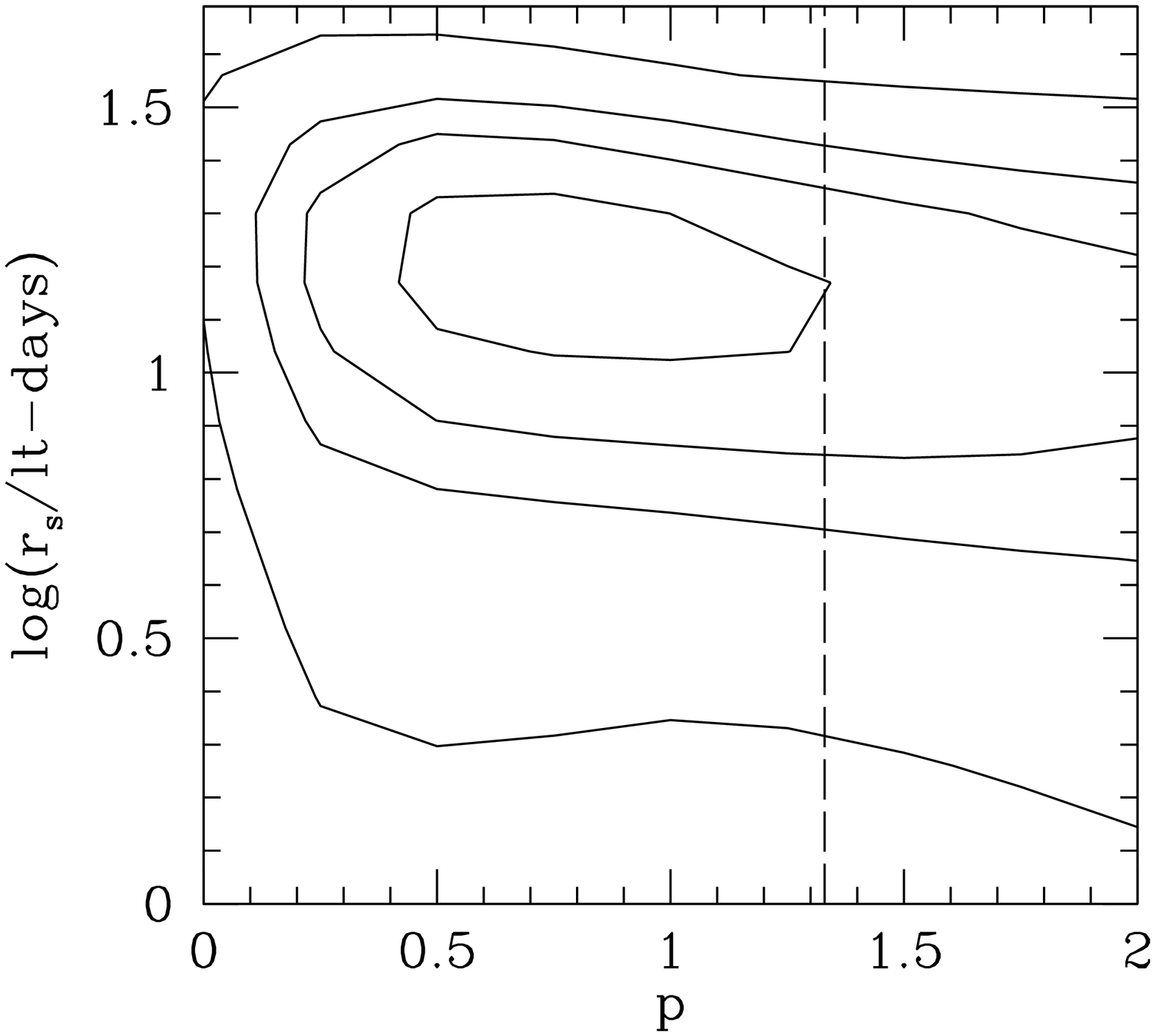}
\caption{\footnotesize{Two-dimensional PDF obtained using the measured chromatic microlensing for HE0435-1223 (Table \ref{map_he0435}) for a logarithmic grid in  $r_s$ (see text).  
Contours correspond to $0.5 \sigma$, $1 \sigma$, $1.5 \sigma$, and $2 \sigma$ respectively.  
The {\em dashed line} corresponds to the value predicted by  the thin disk model ($p=4/3$). }
\label{he0435size}}
\end{center}
\end{figure*}

Following the method described in \S 3 we will use these wavelength
dependent microlensing measurements to estimate the size and
temperature profile of the accretion disk. To make the problem
manageable, we will consider that the global behavior of the continuum
can be defined by two straight lines, one independent of lambda to
describe the data redder than \ion{Mg}{2}, and the other one following the
almost linear dependence of microlensing with $\lambda ^{-1}$ from
\ion{Mg}{2} towards the blue (see Figure \ref{diff_he0435b}). We will take
three points corresponding in wavelength to the F160W band, 8100\AA\ (from
spectroscopic continuum), and u' band (see Table \ref{map_he0435}) to
describe the global dependence of microlensing with wavelength in the
simplest possible way. In Figure \ref{he0435size} we show the 2D
joint PDF of $r_s$ and $p$ conditioned to these three microlensing
measurements. The resulting estimates are $r_s=13^{+5}_{-4}$
light-days at $\lambda_0=1310$~\AA \ and $p=1.2\pm0.6$ ($1\sigma$ level), using uniform 
logarithmic and linear priors for the size and $p$ respectively. The estimated
size ($r_s=7^{+3}_{-2}$~light-days or $r_{1/2}=9^{+4}_{-3}$~light-days, scaled to $0.3M_{\odot}$) is in agreement with the average size estimated by \cite{jimenez14} ($r_s=4.8^{+6.2}_{-2.7}$~light-days at 1026~\AA), and it is consistent with the results in \cite{blackburne11} ($\log (r_{1/2}/\rm cm)=16.09\pm0.19$ or $4.8^{+2.6}_{-1.7}$~light-days at 1208~\AA). This is also consistent with sizes predicted by the black hole-mass size correlation \cite[][$r_s=2.3^{+0.8}_{-0.5}$ at 2500~\AA \, for $M_{BH}=10^9M_{\odot}$]{morgan10}.  
However, this size is large compared to predictions based on flux variations \citep[][$r_s=0.76\times10^{15} \rm cm$ or 0.3~light-days at 3027~\AA]{mosquera11b}, and recent results using R-band light-curves \citep[][$\log (r_s/\rm cm)=15.23^{+0.34}_{-0.33}$ or $0.7^{+0.8}_{-0.4}$~light-days at 3027~\AA]{blackburne14}. 
Our results for $p$ are
 in agreement with previous results (ranging from $0.55\pm0.49$ in
\cite{blackburne11} to $1.3\pm0.6$ in \cite{jimenez14}). Notice that
these results match those of \cite{jimenez14} although
those were estimated using pair $B-A$.

We follow the same procedure \citep{guerras13} to estimate the size of
the \ion{C}{4} emitting region affected by microlensing. In this case, we
only calculate the size of the region at the \ion{C}{4} wavelength. We
obtained $r_s=10^{+15}_{-7}$~light-days. This size is in agreement
with that of the region emitting the blue continuum, indicating
that the microlensed \ion{C}{4} emission likely arises close to the
accretion disk.

\begin{deluxetable}{rc}
\tablecolumns{2}
\tablewidth{0pt} 
\tablecaption{HE0435-1223 chromatic microlensing \label{map_he0435}}
\tablehead{ 
  \colhead{$\lambda_c$ (\AA)} &
  \colhead{$\Delta m_C - \Delta m_L$\tablenotemark{a} (mag)} 
}
\startdata                  
3522    & $0.39\pm0.1$ \\ 
8100    & $0.12\pm0.1$ \\ 
15500   & $0.11\pm0.1$ \\ 
\enddata
\tablenotetext{a}{Difference between the magnitude difference in the continuum and in 
the emission line cores $(m_D-m_B)_C-(m_D-m_B)_L$ for 
MMT and VLT continuum data and including the CASTLES F160W band data (see text).}
\end{deluxetable}

\subsection{WFI2033-4723} \label{wfi2033}

This is a quadruply imaged quasar ($z_S=1.664\pm0.002$ by our
estimation) discovered by \cite{morgan04} lensed by a galaxy at
$z_L=0.66$ \citep{ofek2006,eigenbrod2008}. We present new spectra for
components $B$ and $C$ which are separated
$2\farcs1$. \cite{vuissoz08} found a time delay of $\Delta t_{BC}\sim
62.6^{+4.1}_{-2.3}$ days, in agreement with values estimated by
\cite{morgan05}.

\begin{figure*}
\begin{center}
\includegraphics[width=5cm]{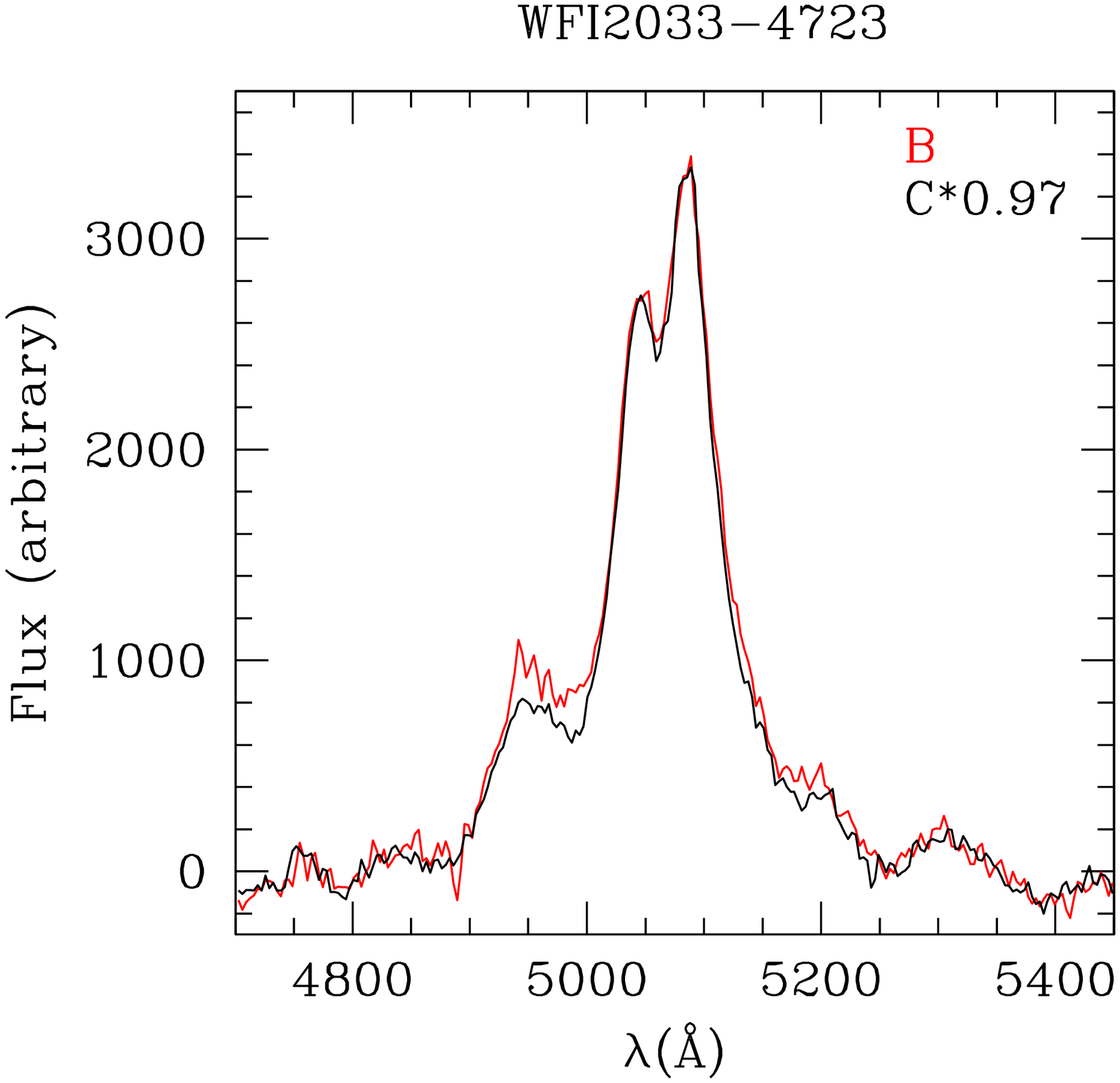}
\includegraphics[width=5cm]{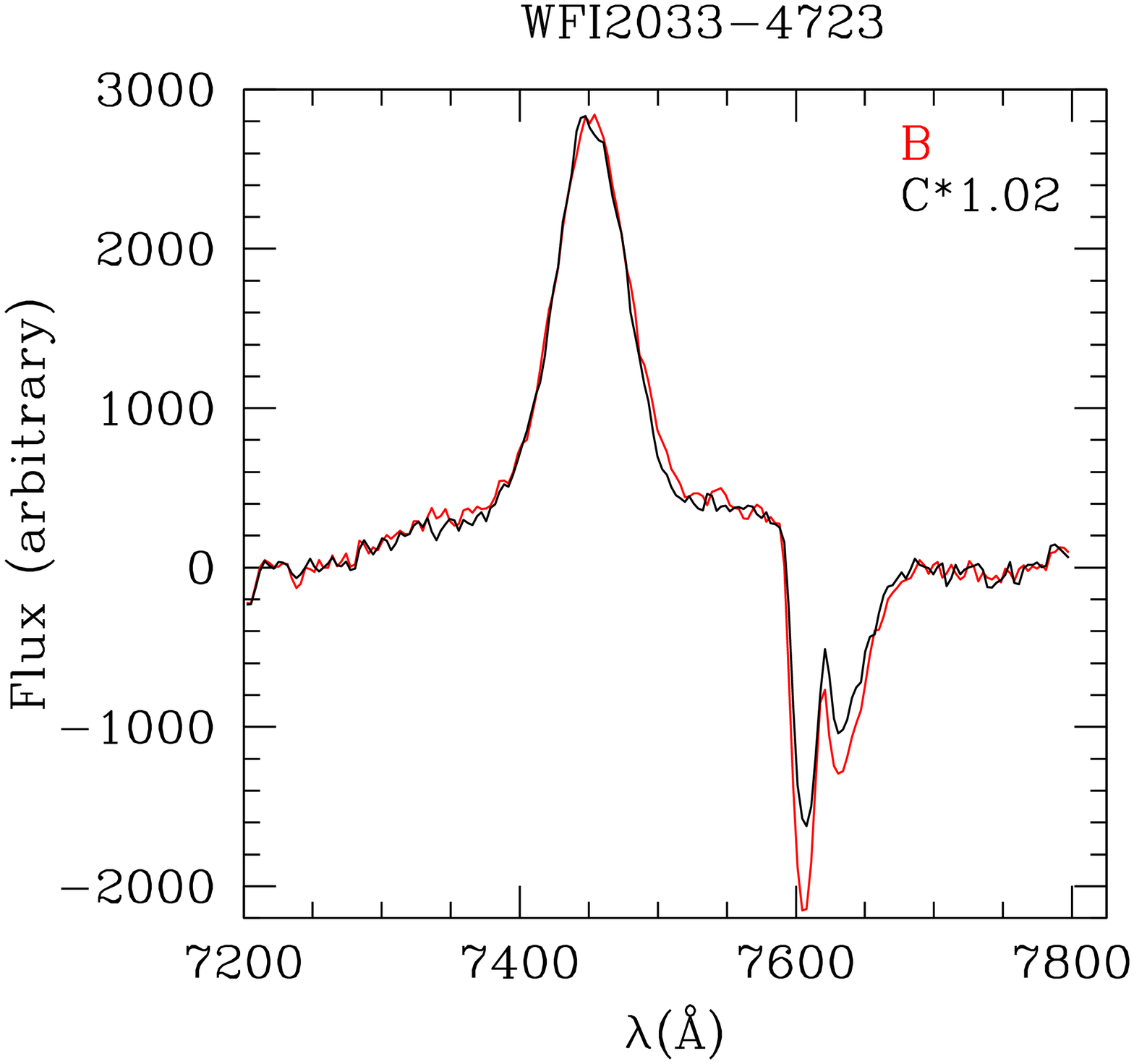}\\
\includegraphics[width=5cm]{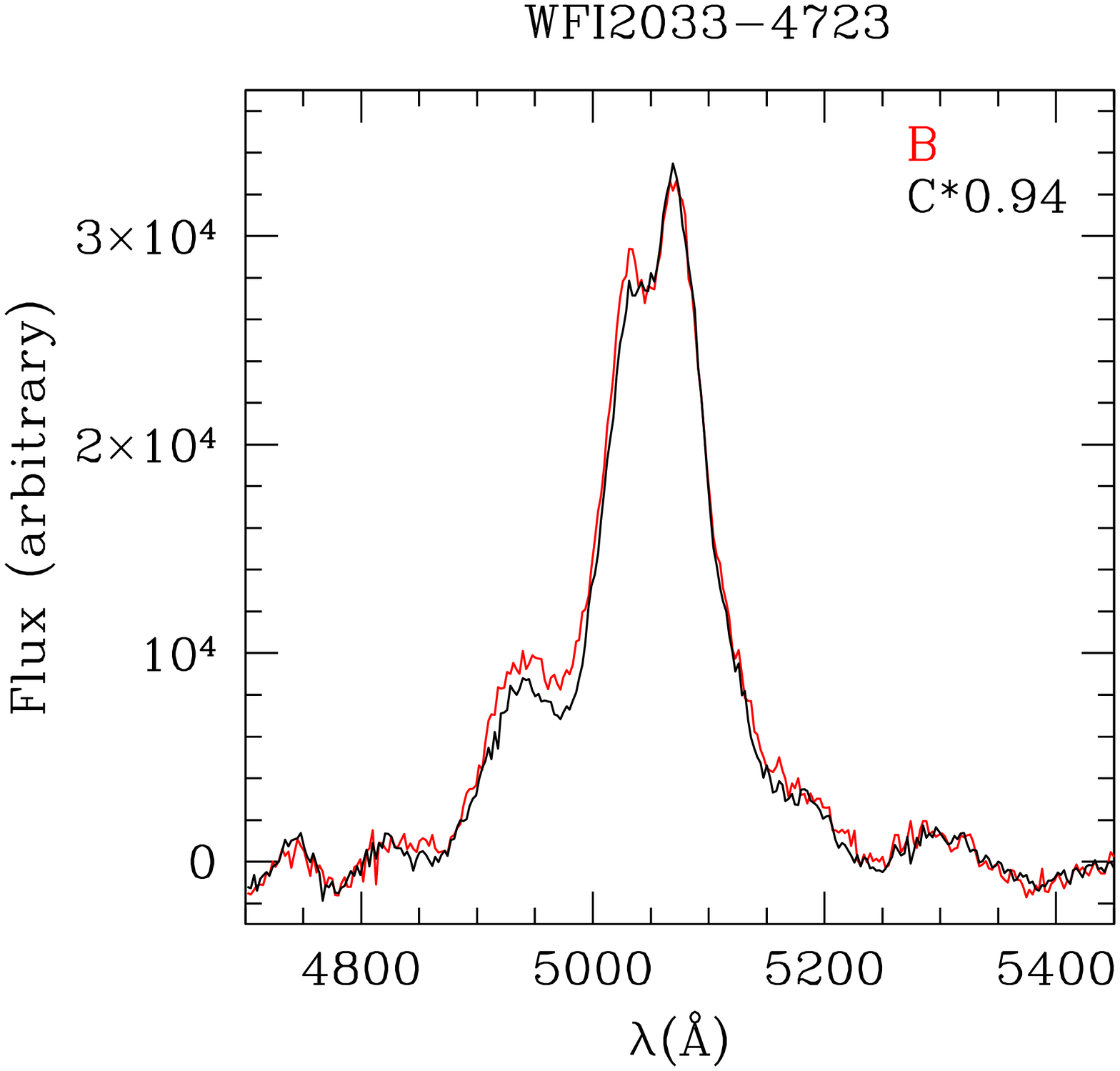}
\includegraphics[width=5cm]{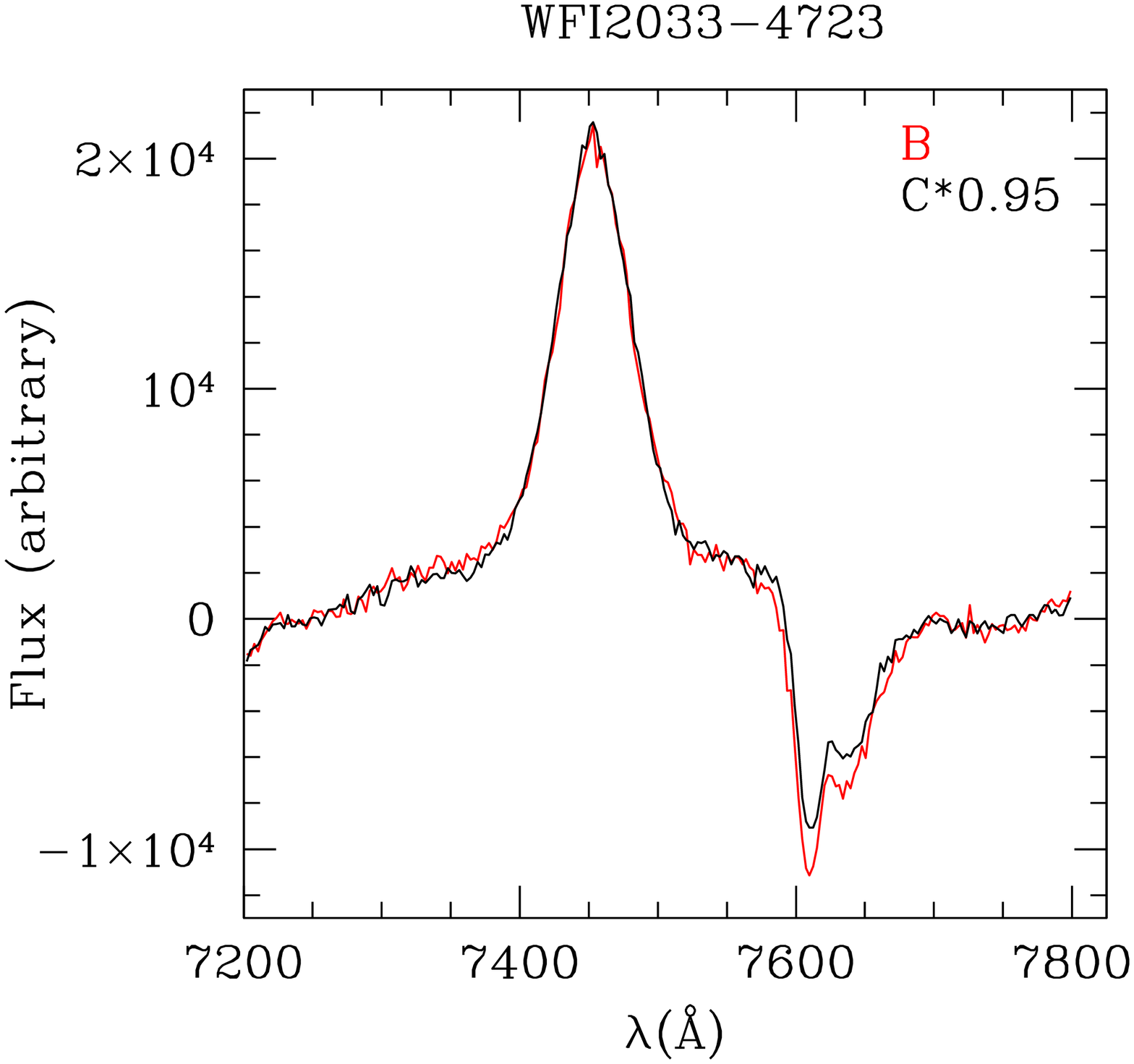}
\caption{\ion{C}{3]}, and \ion{Mg}{2} emission line profile for WFI2033-4723 vs. observed $\lambda$ for 
our VLT data ({\em top}) and VLT deconvolved spectra by \cite{sluse12} ({\em bottom}).
The {\em red line} represents the continuum subtracted emission lines for $B$. 
The {\em black line} represents the continuum subtracted emission line for $C$ 
multiplied by a factor to match the peak of $B$. The factors are shown in each panel. 
\label{prof_wfi2033}}
\end{center}
\end{figure*}

In Figure \ref{prof_wfi2033} we plot the continuum subtracted spectra
for images $B$ and $C$ in the regions corresponding to the \ion{C}{3]} and
  \ion{Mg}{2} emission lines. The spectra have been normalized
  to match the profiles in the region of the core of the line (see \S
  3).  A very interesting result is the significant enhancement (of the $B$
  image compared to the $C$ image) of a bump-like feature present in the
  blue wing of the \ion{C}{3]} line, coincident with an \ion{Al}{3} emission line
    typical of quasar spectra.  Careful inspection of our spectra
    reveals that there is another bump-like feature in the red wing of
    the \ion{C}{3]} line that also appears to be brighter in $B$ than in
      $C$.

At a much lower intensity, two microlensed bumps with a
separation of about $\sim200$~\AA\ are also noticeable at both
sides of the \ion{Mg}{2} emission lines. In the \cite{morgan04} spectra, the
\ion{C}{3]} emission line profiles also show a relative enhancement of the $B$
  image with respect to $C$. This was interpreted as evidence of
  microlensing by \cite{sluse12}, who indicated that the excess might
  also be due to microlensing of the \ion{Al}{3} or \ion{Si}{3]} lines blended
    with the blue wing of \ion{C}{3]}. Unfortunately, the S/N spectra of the
      red side of \ion{C}{3]} and of the \ion{C}{4} lines are not sufficient
        to look for further evidence of these features.

\begin{deluxetable}{llcr}
\tablecolumns{5}
\tablewidth{0pt} 
\tablecaption{WFI2033-4723 magnitude differences \label{mag_wfi2033}}
\tablehead{ 
  \colhead{Region} &
  \colhead{$\lambda_c$ (\AA)} &
  \colhead{Window\tablenotemark{b} (\AA)} &
  \colhead{$m_C-m_B$\tablenotemark{a} (mag)} 
}
\startdata
Continuum & 5140     & 4700-5450 & $0.30\pm0.01$ \\
          & 7560     & 7200-7800 & $0.30\pm0.01$ \\
\hline             
Line  & \ion{C}{3]}$\lambda$1909 & 5060-5100 & $-0.02\pm0.01$   \\
      & \ion{C}{3]} blue wing    & 4930-5000 & $0.17\pm0.02$   \\
      & \ion{C}{3]} red wing     & 5160-5250 & $0.22\pm0.03$   \\
      & \ion{Mg}{2}$\lambda$2800  & 7430-7470 & $0.03\pm0.01$   \\
\enddata
\tablenotetext{a}{VLT data}
\tablenotetext{b}{Integration window.}
\end{deluxetable}

Integrating the bump excess corresponding to image $B$ in the
$[4930,5000]$~\AA \, and $[5160,5250]$~\AA \, ranges, we obtain the
microlensing magnification associated to the region generating the
blue/red wing as \cite[see][]{guerras13}: $\Delta
m^{blue,red}_{CB}=(m_C-m_B)_{blue,red}-(m_C-m_B)_{core}$. We obtain
consistent microlensing estimates for both bumps: $\Delta
m^{blue}=0.17\pm0.02$~mag and $\Delta m^{red}=0.22\pm0.03$~mag.  The
presence of the two bumps with similar enhancements on both sides of
the line suggest a common origin for both structures (as an
alternative to a microlensed \ion{Al}{3} line).  In several AGNs
\cite[see][and references therein]{popovic04} the presence of these
features in the wings of a BEL, which remind us of the double-peak
characteristic of kinematics confined in a plane, have been
interpreted as evidence of the accretion disk surrounding the massive
central black-hole. Both elements, the separation between the bumps
$\sim250$\AA\ ($\sim 15000\rm\, km\,s^{-1}$) and the fact that only
the bumps are microlensed, would be in agreement with the hypothesis
that these features arise from an inner, compact region of the BLR,
likely residing on the accretion disk. An alternative possibility is
that the bumps arise from a region with biconical geometry
\citep{abajas07}.

\begin{deluxetable}{rc}
\tabletypesize{\scriptsize}
\tablecolumns{2}
\tablewidth{0pt} 
\tablecaption{WFI2033-4723 chromatic microlensing \label{map_wfi2033}}
\tablehead{ 
  \colhead{$\lambda_c$ (\AA)} &
  \colhead{$\Delta m_C - \Delta m_L$\tablenotemark{a} (mag)} 
}
\startdata                  
 4300   & $0.29\pm0.04$ \\ 
 5500   & $0.31\pm0.01$ \\ 
 8700   & $0.34\pm0.01$ \\ 
\hline	   	       	    
 5439   & $0.31\pm0.04$ \\ 
 8012   & $0.19 \pm0.1$ \\ 
15500   & $0.07\pm0.03$ \\ 
\hline	   	       	    
 3522   & $0.28\pm0.01$ \\ 
 5500   & $0.17\pm0.01$ \\ 
 8500   & $0.11\pm0.01$ \\ 
\hline	   	       	    
 3522   & $0.70\pm0.03$ \\ 
 9114   & $0.26\pm0.02$ \\ 
16500   & $0.13\pm0.03$ \\ 
\enddata
\tablenotetext{a}{Difference between the magnitude difference in the continuum and in the emission line cores $(m_C-m_B)_C-(m_C-m_B)_L$ for a) VLT data, b) CASTLES data,  c) re-analysis of \cite{sluse12} deconvolved spectra + \cite{morgan04} + \cite{vuissoz08} (see text), and d) \cite{blackburne11} data.}
\end{deluxetable}

\begin{figure*}
\begin{center}
\plotone{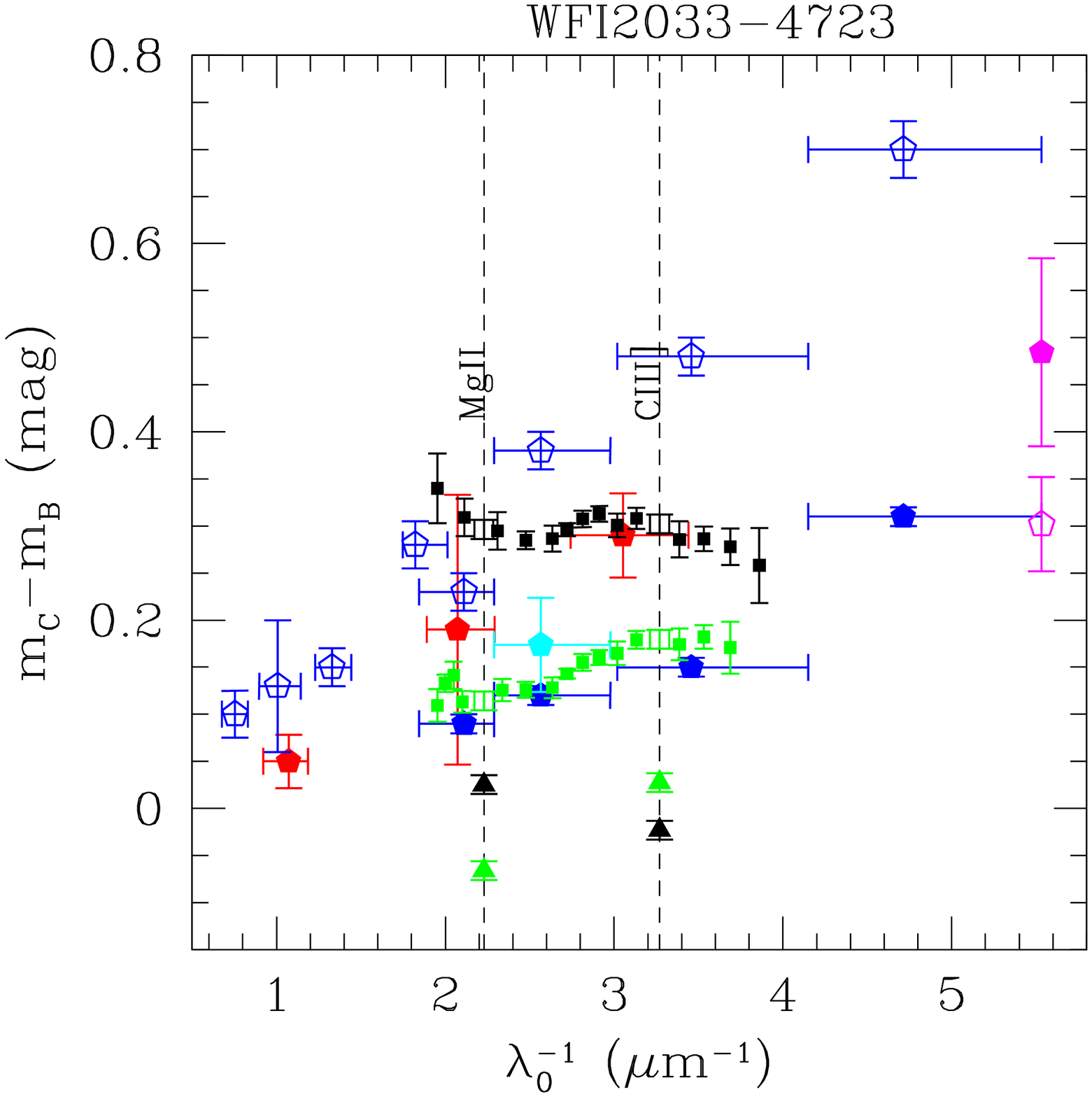}
\caption{\footnotesize{Magnitude differences $m_C-m_B$ vs $\lambda_0^{-1}$ ($\lambda$ in the lens galaxy restframe) for  WFI2033-4723.
Symbols represent the integrated continuum obtained from
(broad-band) CASTLES ({\em solid red pentagon}), \cite{morgan04} ({\em solid blue pentagon}),
\cite{sluse12} ({\em solid green square}), \cite{vuissoz08} ({\em cyan pentagon}), and \cite{blackburne11} ({\em open blue pentagon}). X-ray data obtained by \cite{pooley07} ({\em open magenta pentagon}), and \cite{pooley12} ({\em solid magenta pentagon}) are plotted arbitrarily at 3000 \AA\, for comparison.  The
     {\em solid squares} represent the magnitude differences from the
     integrated continuum in our spectra ({\em black}) and \cite{sluse12} spectra ({\em green}), and the {\em open squares} 
     integrated fitted continuum under the emission lines ({\em black}, {\em green})
     respectively.  {\em Triangles} are the magnitude difference
     in the emission line core for our data ({\em solid black}) and \cite{sluse12} data ({\em solid green}) respectively.  }
\label{diff_wfi2033a}}
\end{center}
\end{figure*}

In Figure \ref{diff_wfi2033a} (Table \ref{mag_wfi2033}) we present the emission line and
continuum ratios for our data and for other data in the
literature. The emission line ratios from \cite{sluse12} and from our
data agree within uncertainties and show no evidence of
chromaticity. We obtain an average value for the emission lines ratio
of $\Delta m_L=-0.01\pm0.02$~mag (we have not considered the
discrepant emission line ratios from \cite{morgan04}, because they
have been obtained integrating in a window of 200~\AA).

\begin{figure*}
\begin{center}
\plotone{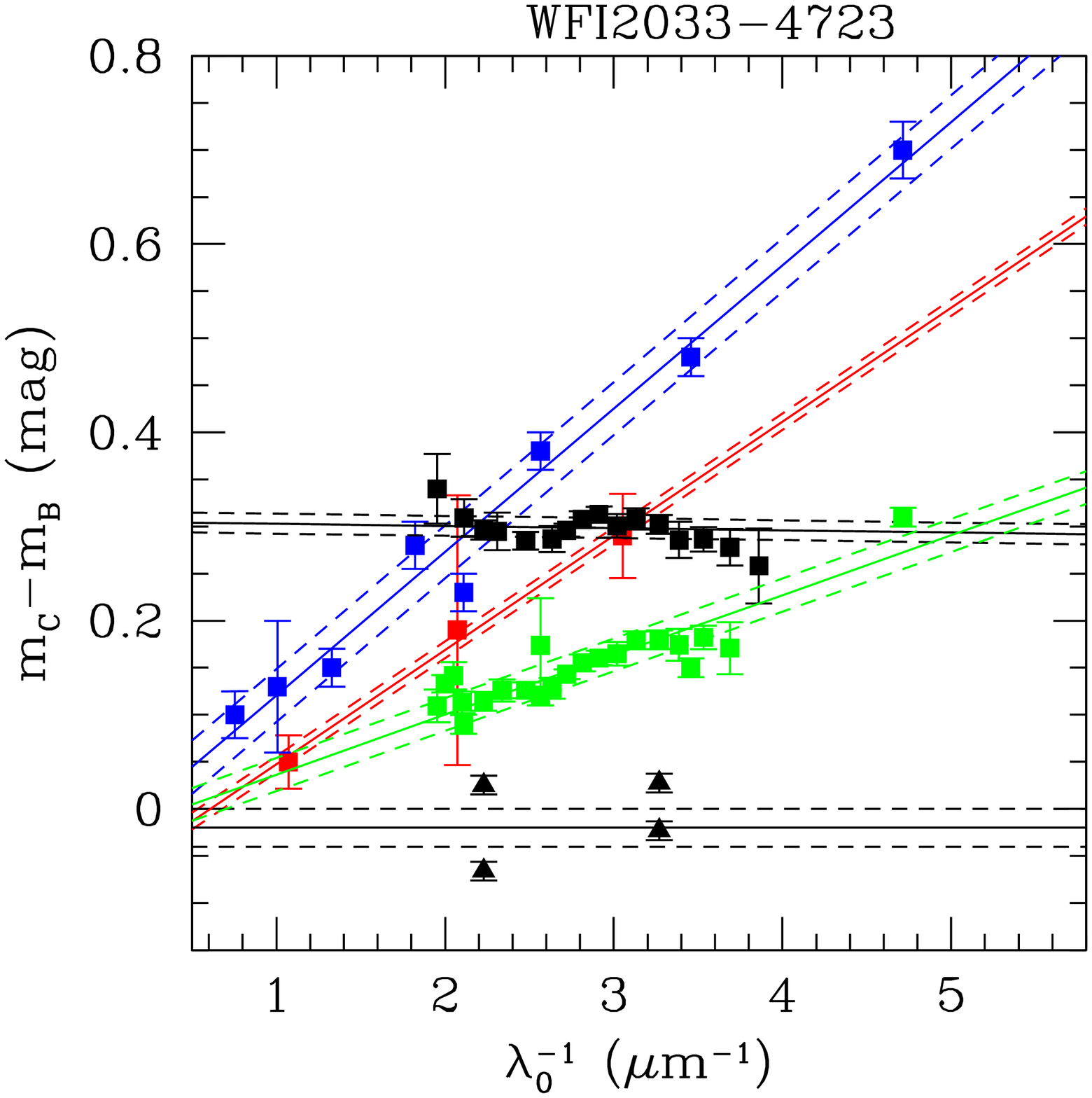}
\caption{Models fitted to the data corresponding to four different
  epochs shown in Figure \ref{diff_wfi2033a} (see text).  {\em
    Squares} and {\em triangles} represent continuum (2004 {\em solid red squares}; 2003-2005 {\em solid green squares}; 2007-2008 {\em solid blue squares}) and line core
  data respectively ({\em black solid triangle}).  {\em Solid lines} represent the function fitted
  to the continua and the average of the emission lines.  {\em Dashed
    lines} are the standard deviations for the continuum fits and the
  standard error of the mean for the emission lines.
\label{diff_wfi2033b}}
\end{center}
\end{figure*}

The continuum ratios, however, show a great change in amplitude at
bluer wavelengths (i.e. chromaticity) showing a varying slope from
2003 to 2008, which can be explained by microlensing magnification of
image $B$ (Figure \ref{diff_wfi2033b}). This trend is better seen if
we group the data around four slopes that correspond to different
epochs or observations (green, red, blue, and black in Figure
\ref{diff_wfi2033b}).
In 2008 the slope of the chromaticity changes, which likely requires a
combination of microlensing in both components to be explained if a
small size for the blue emission is assumed. 

\begin{figure*}
\begin{center}
\includegraphics[width=6cm]{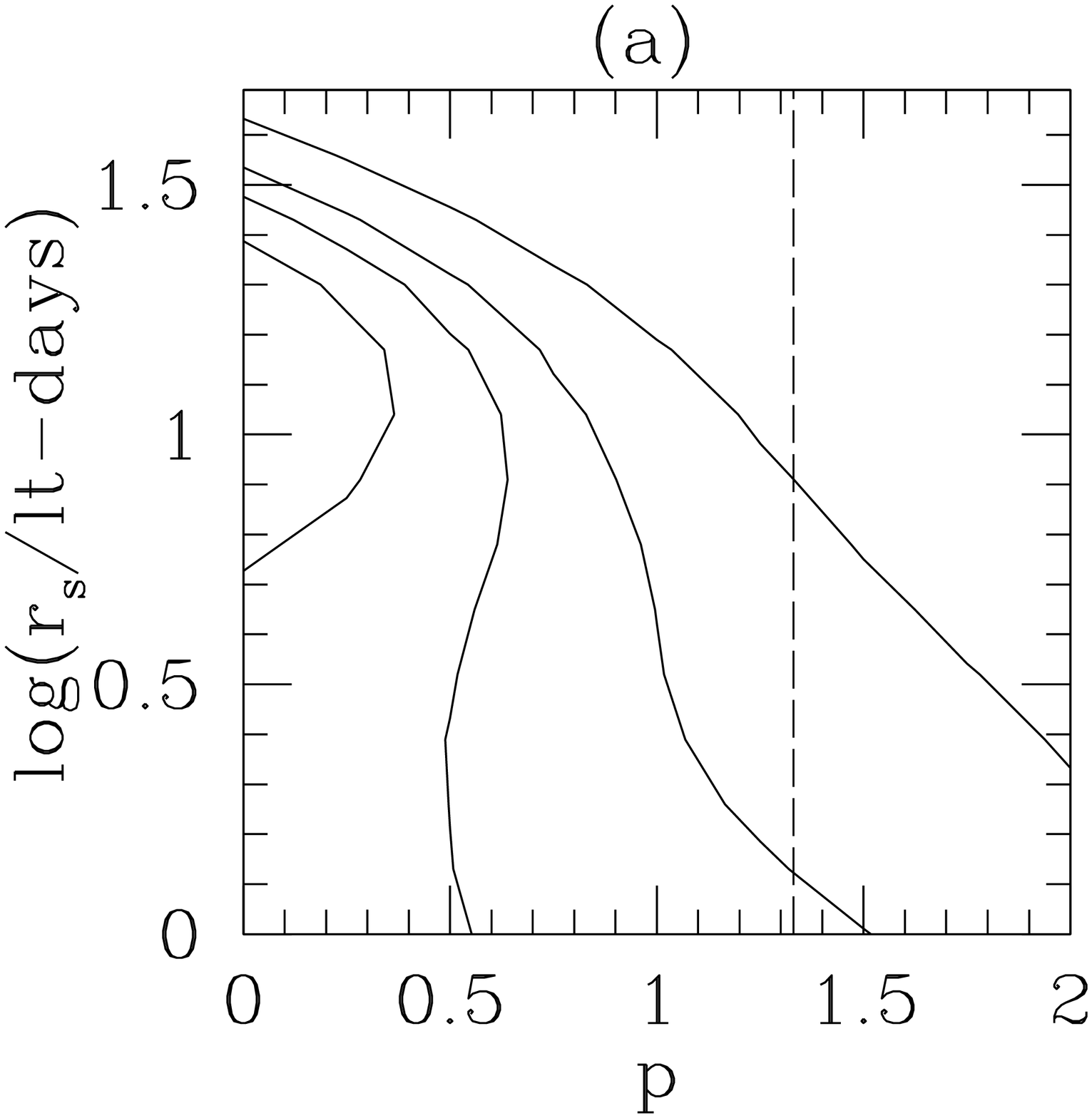}
\includegraphics[width=6cm]{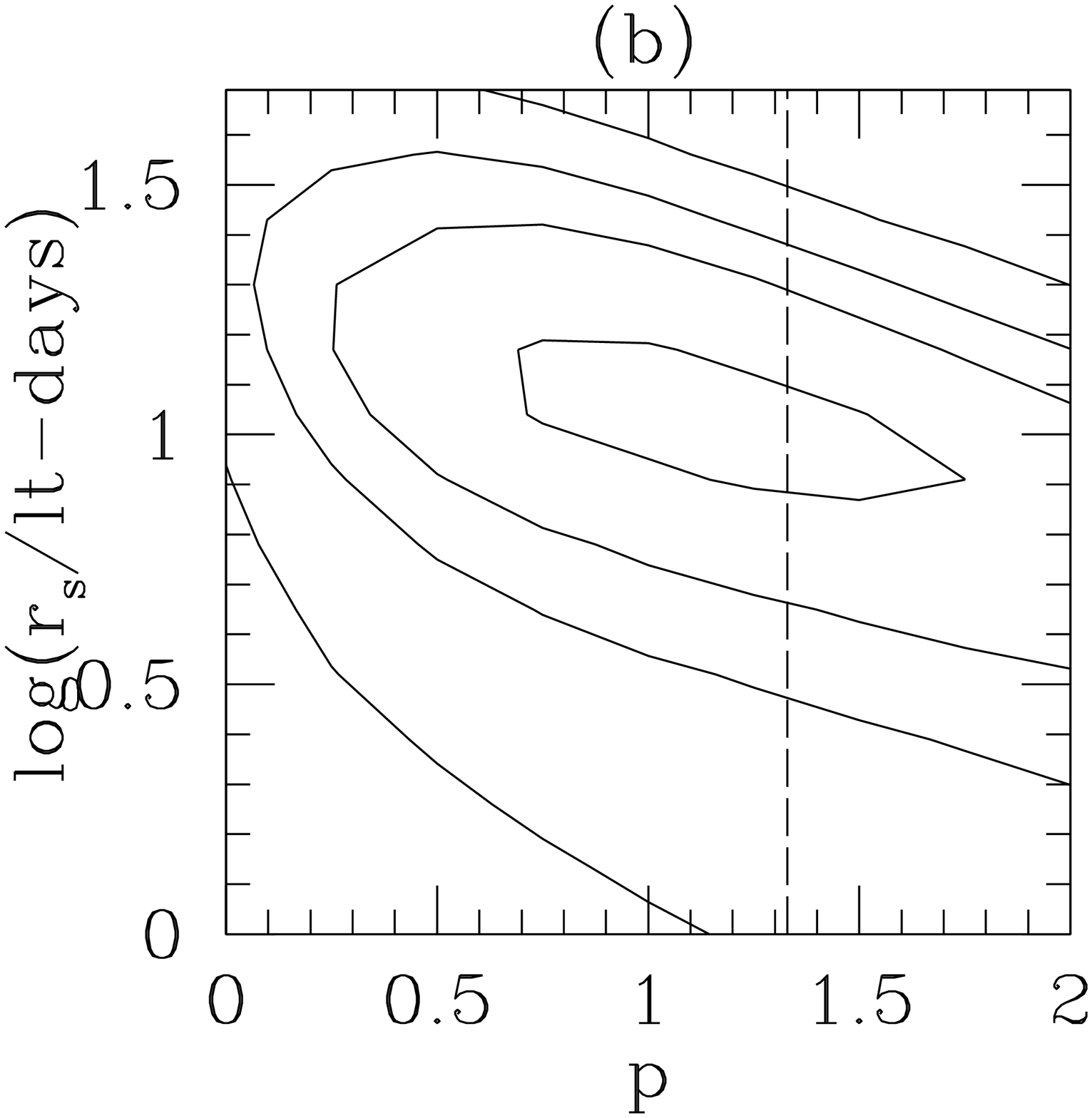}
\includegraphics[width=6cm]{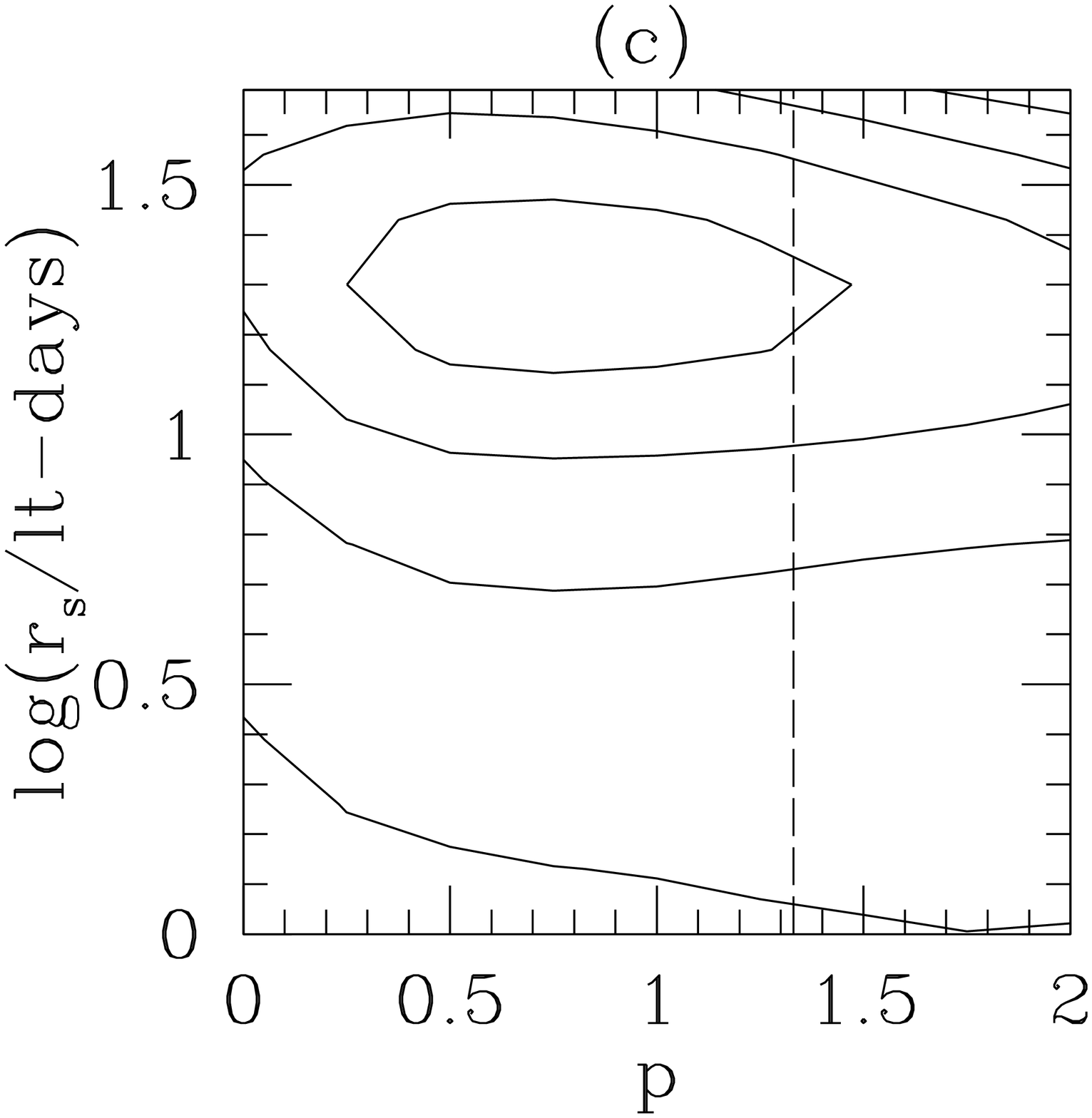} 
\includegraphics[width=6cm]{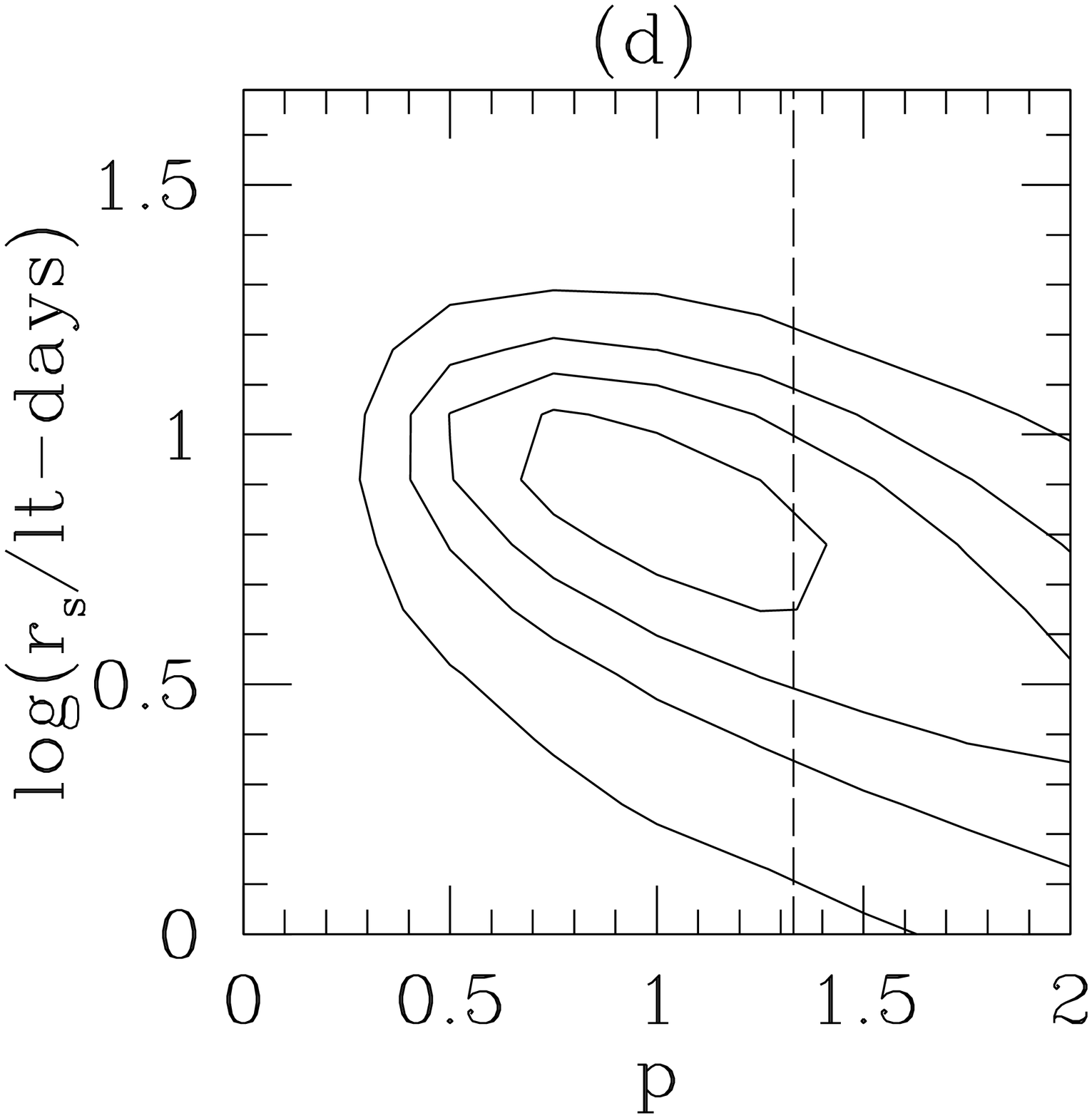}\\
\includegraphics[width=6cm]{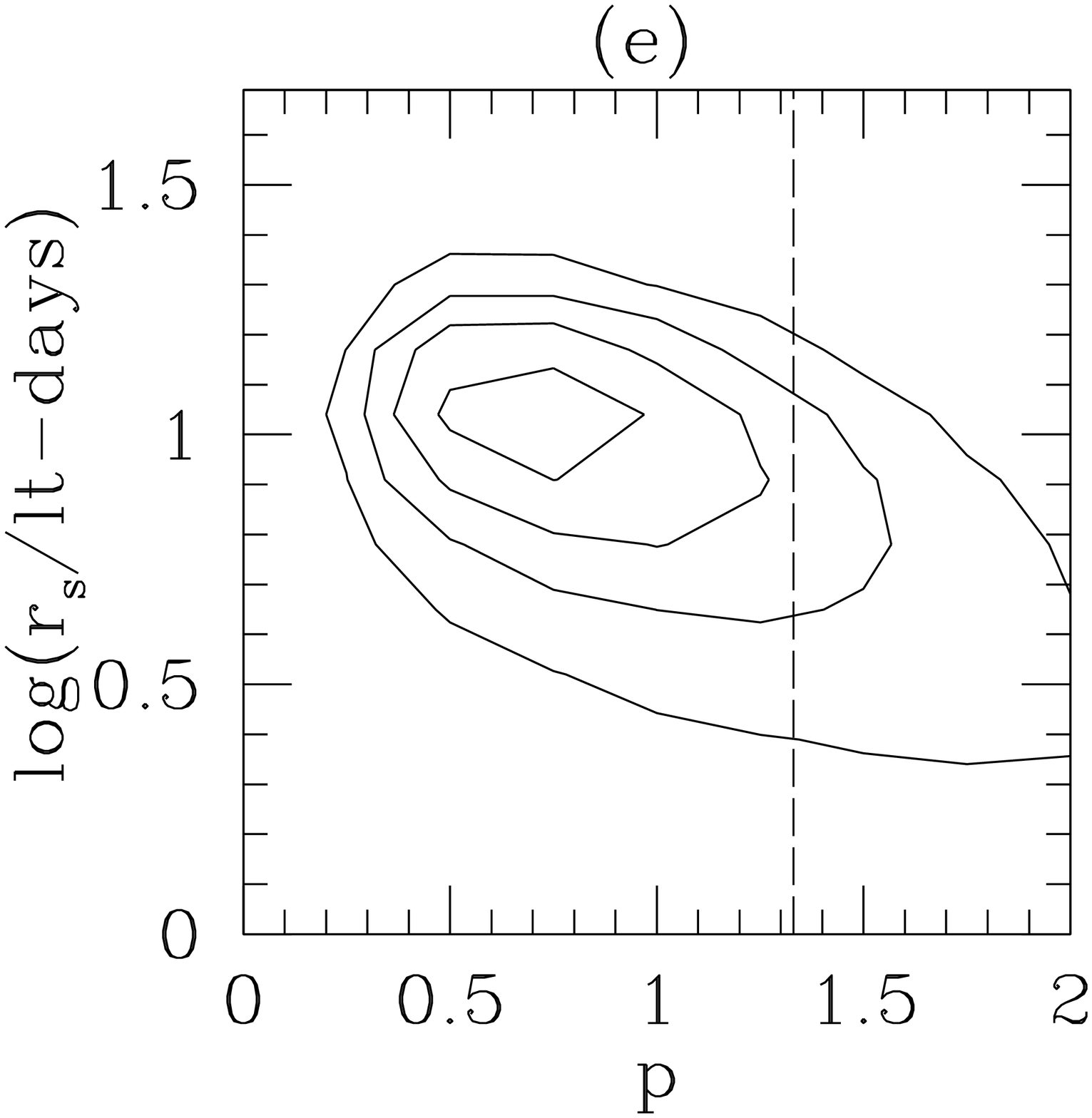}
\caption{\footnotesize{Two-dimensional pdfs obtained using the measured 
chromatic microlensing for WFI2033-4723 (Table \ref{map_wfi2033}) for
 logarithmic grids in $r_s$. (a) VLT, (b) CASTLES, (c) \cite{sluse12},
\cite{morgan04}, and \cite{vuissoz08}, (d) \cite{blackburne11}, (e)
product of the four maps.  Contours correspond to $0.5 \sigma$, $1
\sigma$, $1.5 \sigma$, and $2 \sigma$ respectively.  The {\em dashed
  line} corresponds to the value predicted by the thin disk model
($p=4/3$). }
\label{wfi2033size}}
\end{center}
\end{figure*}

Following the method described in \S 3 we estimate the size, $r_s$,
and the logarithmic slope, $p$, of the size dependence with wavelength
for each one of the 4 epochs described above (Table \ref{mag_wfi2033}). In Figure
\ref{wfi2033size} we present the PDF of $r_s$ and $p$ for each one of
the epochs and the combined PDF. Notice that the slope in our 
continuum data is small (meaning small chromaticity within our error bars, {\em black squares} in Figure \ref{diff_wfi2033b} ), i.e. parameters are largely unconstrained for this epoch (Figure \ref{wfi2033size}~a). 
The resulting estimates for the
combined PDF are $r_s=10^{+3}_{-2}$ light-days (log prior) at
$\lambda_0=1310$~\AA \ and $p=0.8\pm0.2$ (linear prior). 
This disk size 
value ($r_s=6^{+2}_{-1}$~light-days or $r_{1/2}=7^{+2}_{-1}$~light-days, scaled to $0.3M_{\odot}$) is in agreement with the average size estimated by \cite{jimenez14} ($r_s=4.8^{+6.2}_{-2.7}$~light-days at 1026~\AA) and \cite{morgan10} ($r_s=2.3^{+0.8}_{-0.5}$ at 2500~\AA \, for $M_{BH}=10^9M_{\odot}$). 
However, we cannot reconcile our results ($r_s$ and $p$) with those of \cite{blackburne11} ($r_{1/2}=19.8^{+8.2}_{-5.9}$~light-days at 1233~\AA, using 0.3M$_{\odot}$ microlenses, $p=-0.63\pm0.52$). Although these last authors obtain size estimates that decrease with wavelength, they do not rule out positive values for $p$ and suggest that the anomalous flux ratios might be caused by unusual caustic patterns.

Using the same procedure, we estimate a size of $r_s=11^{+28}_{-7}$
light-days for the region generating the emission line bumps on both
sides of the \ion{C}{3]} line. Under the hypothesis of a common origin for
  both bumps, we can take this size as the radius of the accretion
  disk, infer a velocity of 7500$\rm\, km\, s^{-1}$ from the
  separation between the bumps and, assuming Keplerian circular
  rotation, estimate a mass of $1.2^{+3.1}_{-0.8}\times 10^8
  M_{\odot}$ for the supermassive black hole.

\subsection{HE2149-2745} \label{he2149}

HE2149-2745 was discovered by \cite{wisotzki96}; it consists of two
images $A$ and $B$ separated by $1\farcs70$ at
$z_S=2.033\pm0.005$. The lens galaxy is at $z_L=0.603\pm0.001$
\citep{eigenbrod2007}. Image $B$ is separated by $0\farcs34$ from the
main lens galaxy. Chromatic microlensing was detected in spectra taken
by \cite{burud02}. These authors estimated a time delay of $103\pm12$
days.

\begin{figure*}
\begin{center}
\includegraphics[width=5cm]{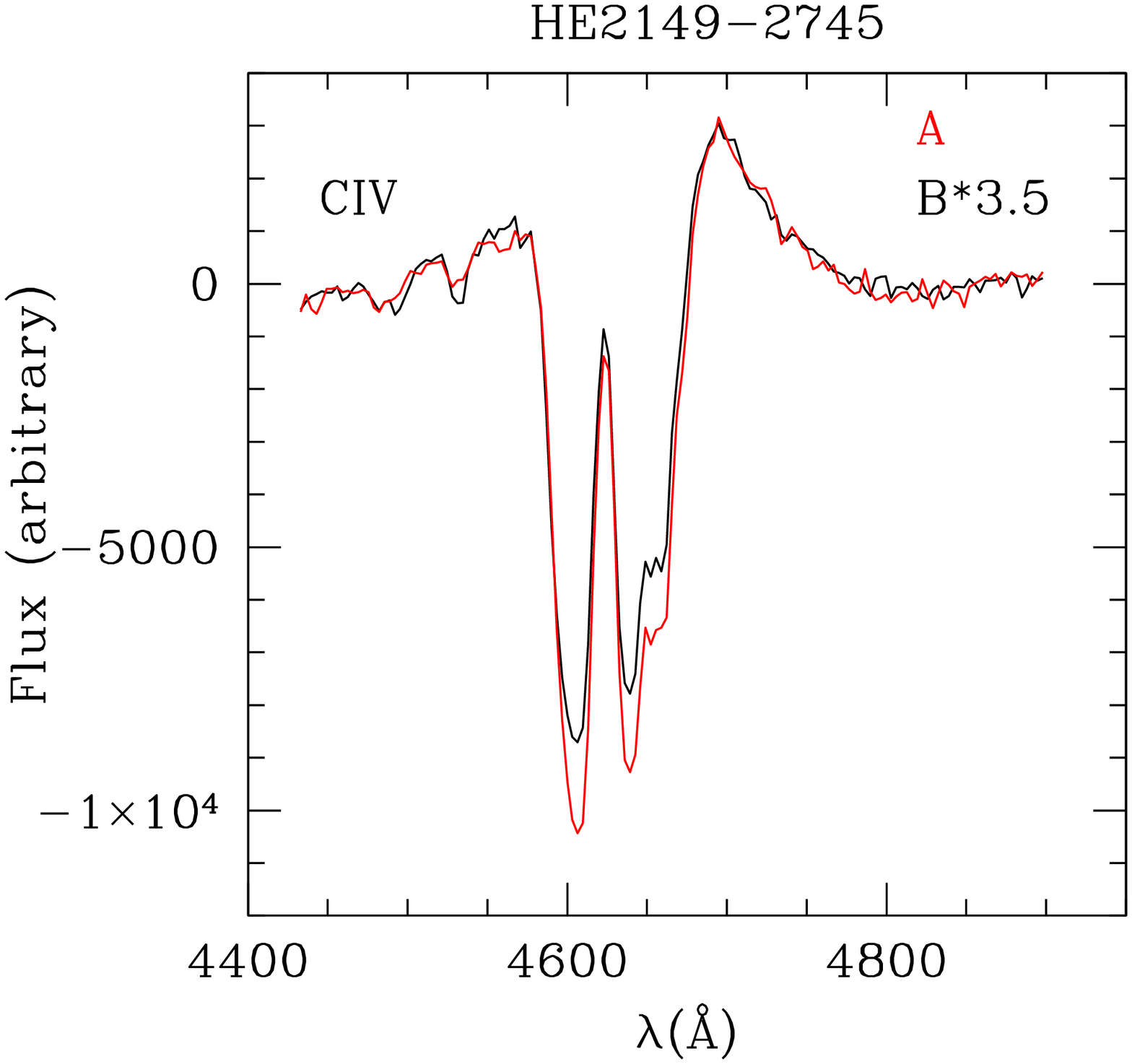}
\includegraphics[width=5cm]{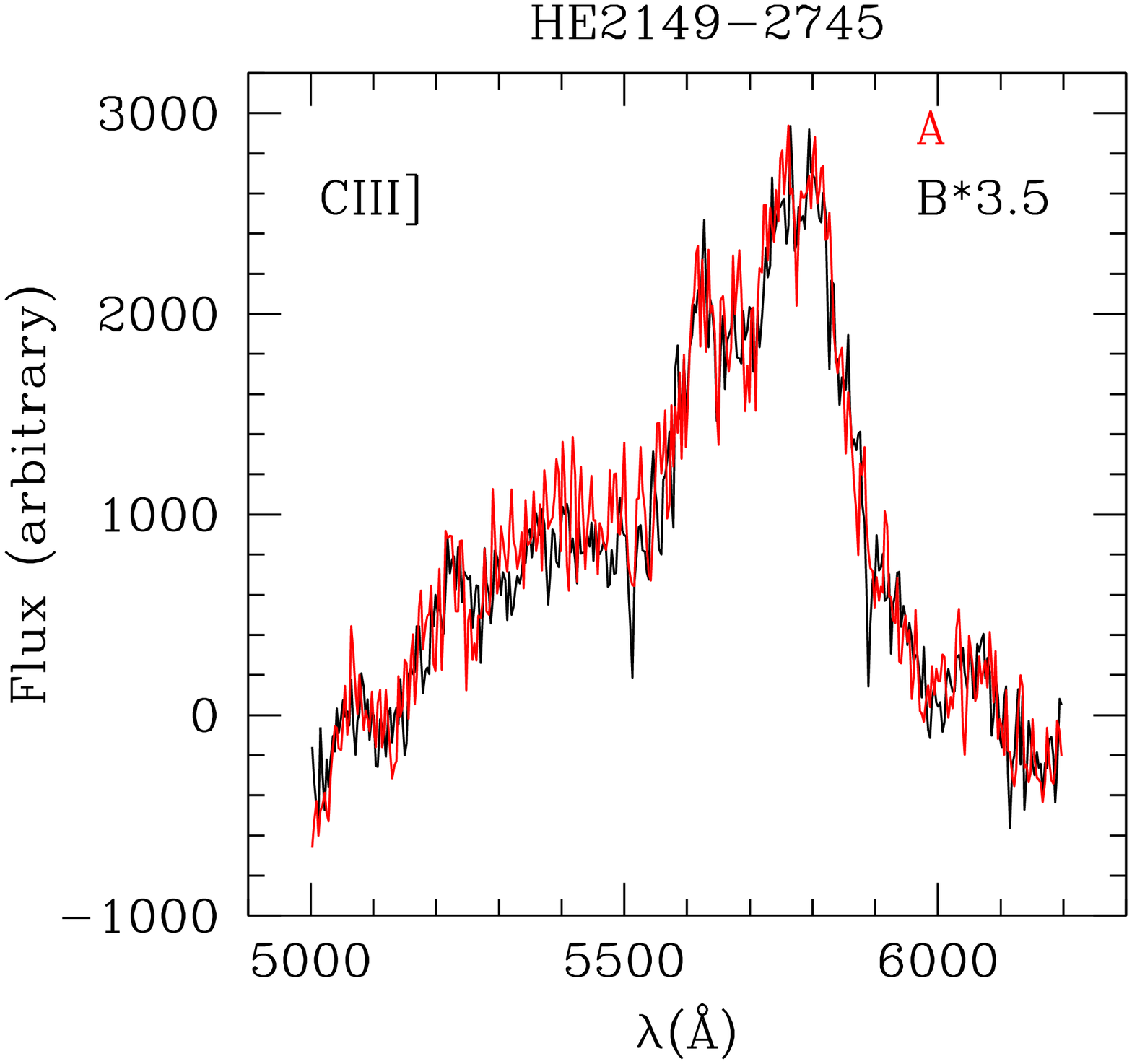}
\includegraphics[width=5cm]{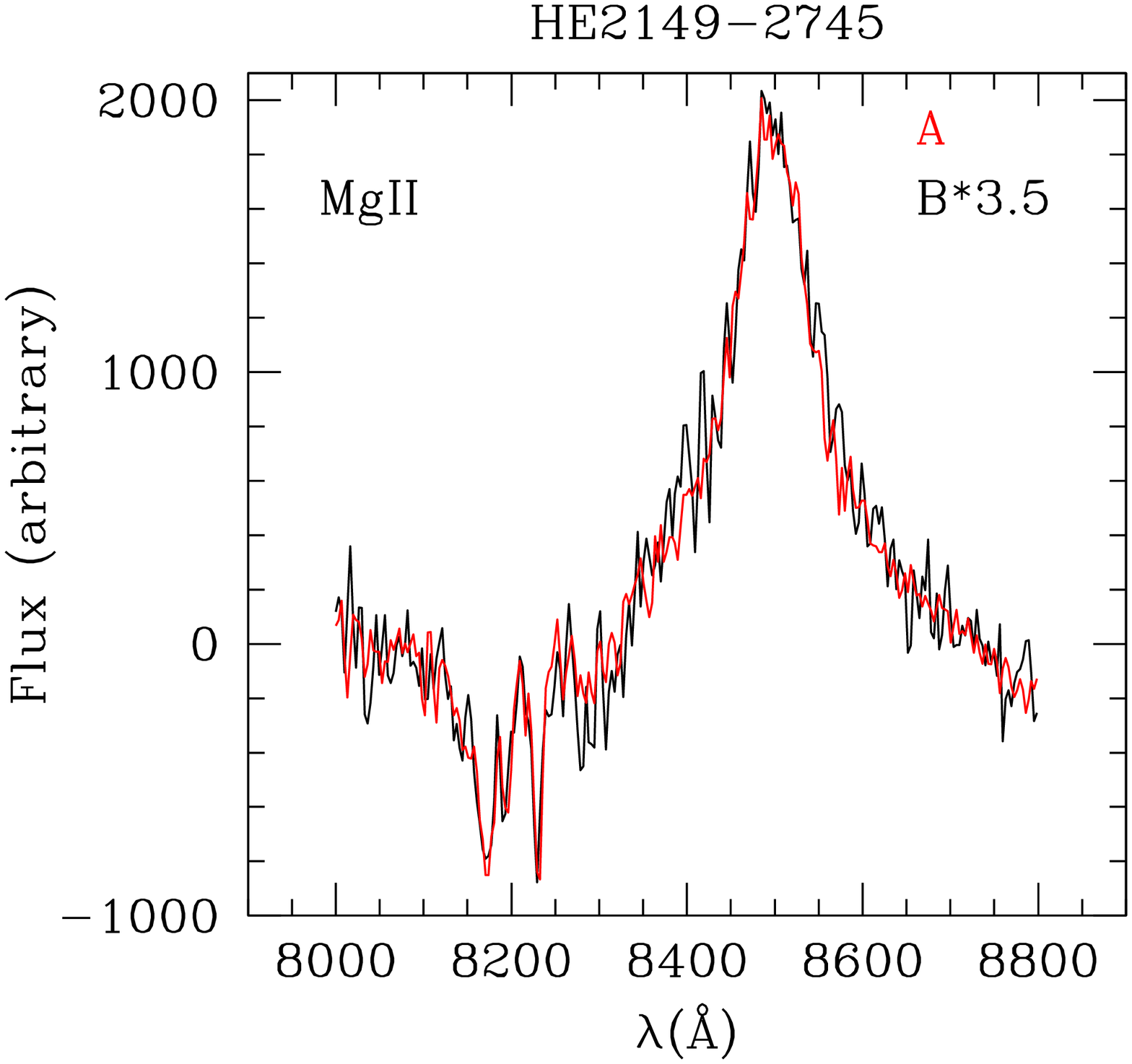}\\
\includegraphics[width=5cm]{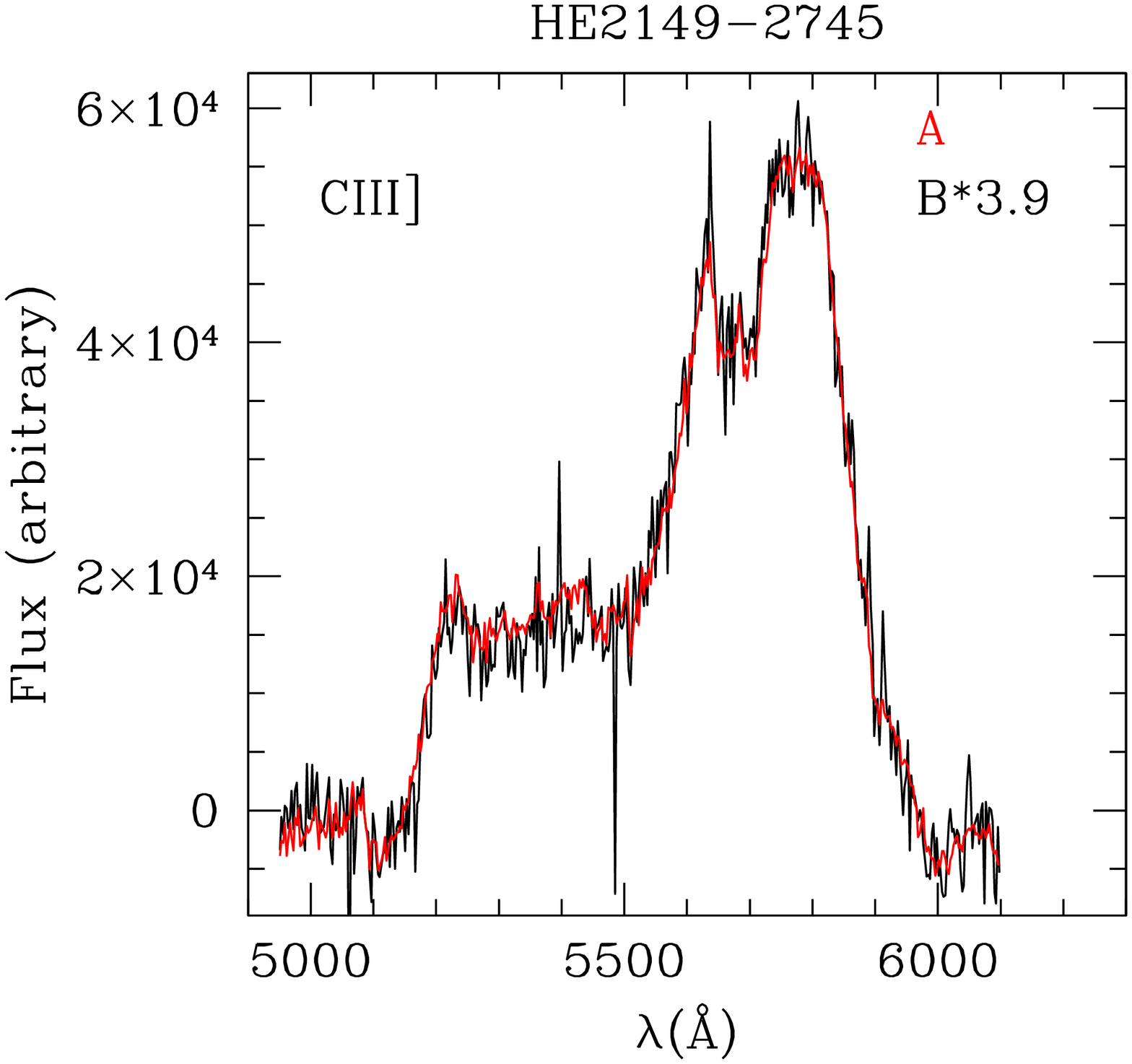}
\includegraphics[width=5cm]{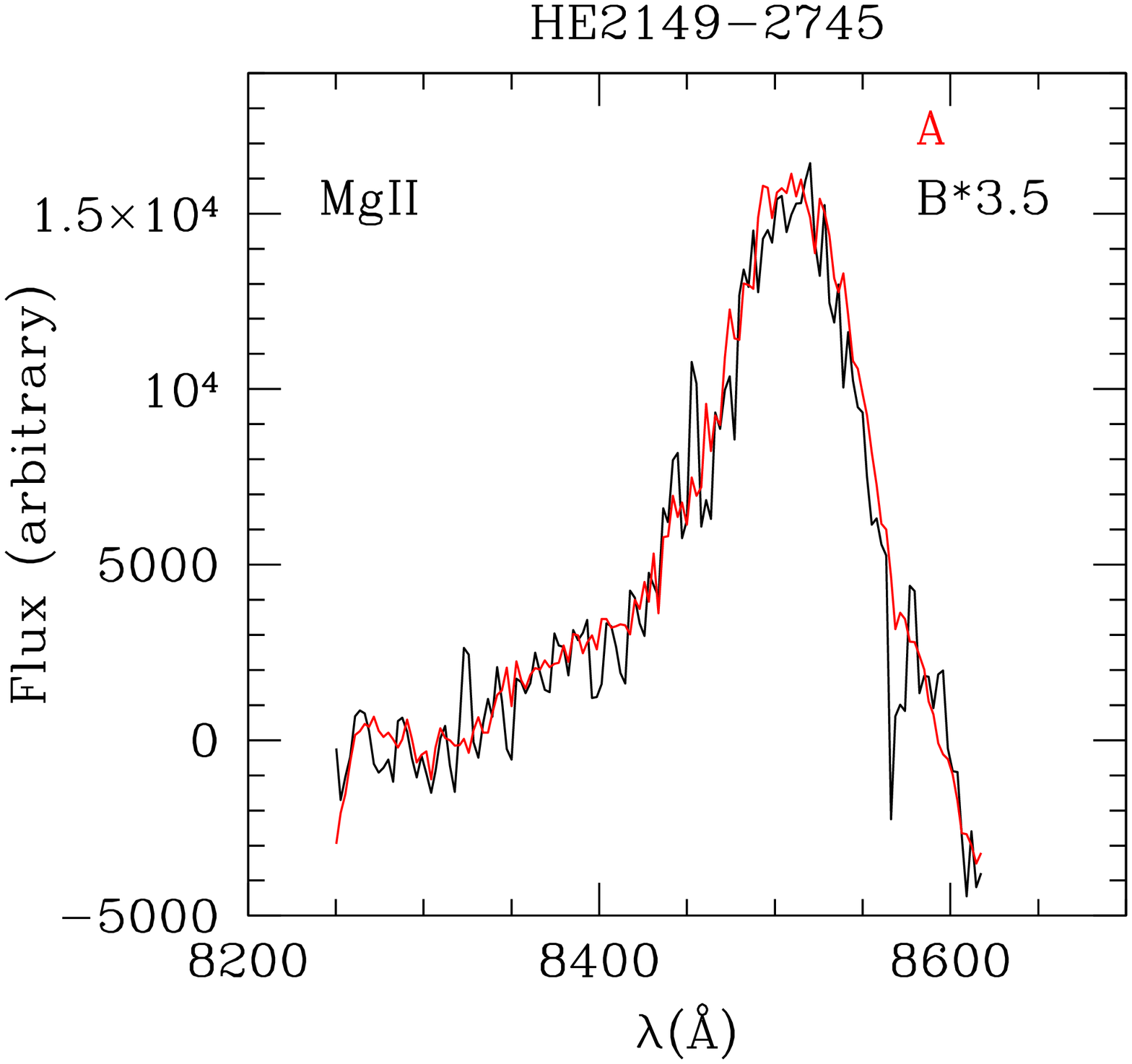}
\caption{\ion{C}{4}, \ion{C}{3]}, and \ion{Mg}{2} emission line profile for HE2149-2745
vs. observed $\lambda$ for VLT data ({\em top}) and deconvolved
spectra by \cite{sluse12} ({\em bottom}).  The {\em red line}
represents the continuum subtracted emission lines for $A$.  The {\em
  black line} represents the continuum subtracted emission line for
$B$ multiplied by a factor to match the peak of $A$. The factors are
shown in each panel.
\label{prof_he2149}}
\end{center}
\end{figure*}

\begin{deluxetable}{llcrr}
\tablecolumns{5}
\tablewidth{0pt} 
\tablecaption{HE2149-2745 magnitude differences \label{mag_he2149}}
\tablehead{ 
  \colhead{Region} &
  \colhead{$\lambda_c$ (\AA)} &
  \colhead{Window\tablenotemark{c} (\AA)} &
  \colhead{$m_B-m_A$\tablenotemark{a} (mag)} &
  \colhead{$m_B-m_A$\tablenotemark{b} (mag)} 
}
\startdata
Continuum & 4170     & 4000-4350 & $-0.11\pm0.02$  & \nodata        \\
          & 5140     & 5000 6200 & $-0.14\pm0.01$  & $-0.12\pm0.02$ \\
          & 7560     & 8250 8650 & $-0.28\pm0.02$  & $-0.22\pm0.01$ \\
\hline             
Line  & \ion{C}{4}$\lambda$1549   & 4170-4195 & $-0.37\pm0.02$ & \nodata         \\
      & \ion{C}{3]}$\lambda$1909 & 5720 5830 & $-0.34\pm0.01$ & $-0.36\pm0.01$   \\
      & \ion{Mg}{2}$\lambda$2800  & 8480 8540 & $-0.47\pm0.02$ & $-0.37\pm0.01$   \\
\enddata
\tablenotetext{a}{VLT data} 
\tablenotetext{b}{\cite{sluse12} }
\tablenotetext{c}{Integration window.}
\end{deluxetable}

In Figure \ref{prof_he2149} we present the continuum subtracted
spectra for the $A$ and $B$ images in the regions corresponding to the
\ion{C}{4}, \ion{C}{3]} and \ion{Mg}{2} emission lines. The spectra match very well, after
  normalization using $A/B=3.5$, except for the absorption in
  \ion{C}{4}. \cite{sluse12} also find that the absorbed fractions of the \ion{C}{4}
  emission line do not agree once scaled and they attribute the
  difference to time-variable broad absorption together with a 
  time-delay. They found a chromatic difference between the spectra
  ($A/B=4$ for CIII and $A/B=3.5$ for \ion{Mg}{2}) which they attribute to
  dust extinction on image $B$ and/or intrinsic variability combined
  with a time delay of $\sim 103$ days.

\begin{figure*}
\begin{center}
\plotone{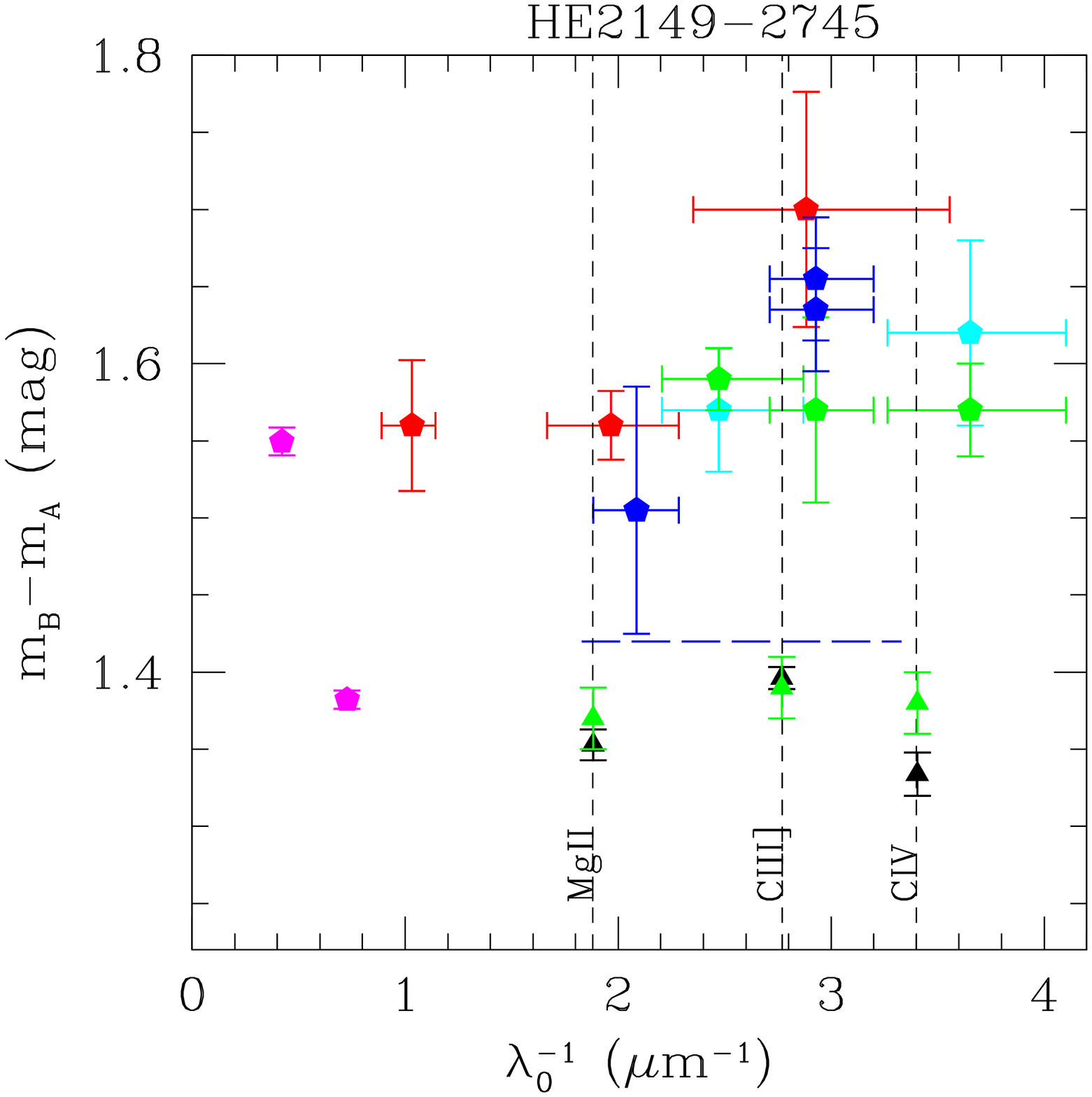}
\caption{\footnotesize{Magnitude differences 
$m_B-m_A$ vs $\lambda_0^{-1}$ ($\lambda$ in the
lens galaxy restframe) for HE2149-2745.
{\em Solid pentagons} represent the integrated continuum obtained from (broad-band): CASTLES ({\em red}), \cite{wisotzki96} ({\em cyan}), \cite{lopez98} ({\em green}), \cite{fadely11} ({\em magenta}), and \cite{burud02} ({\em blue}). 
The integrated continuum in our spectra is contaminated by the lens galaxy and it is not shown here (see text).  
{\em Triangles} are the magnitude difference in the narrow emission line (NEL) for the spectra ({\em black}, {\em green}).} The {\em dashed blue line} represents the relative magnification obtained by \cite{burud02} from spectra. 
\label{diff_he2149a}}
\end{center}
\end{figure*}

Flux comparison between our spectroscopic continua and CASTLE data
shows that there is contamination by the lens galaxy. We estimate
(using F555W and F814W fluxes) that the contamination in the B spectra
could be up to 60\%, while in the deconvolved spectra it could be around
30\%. In the following we will use the broad-band data to obtain the
magnitude difference in the continuum.

\begin{figure*}
\begin{center}
\plotone{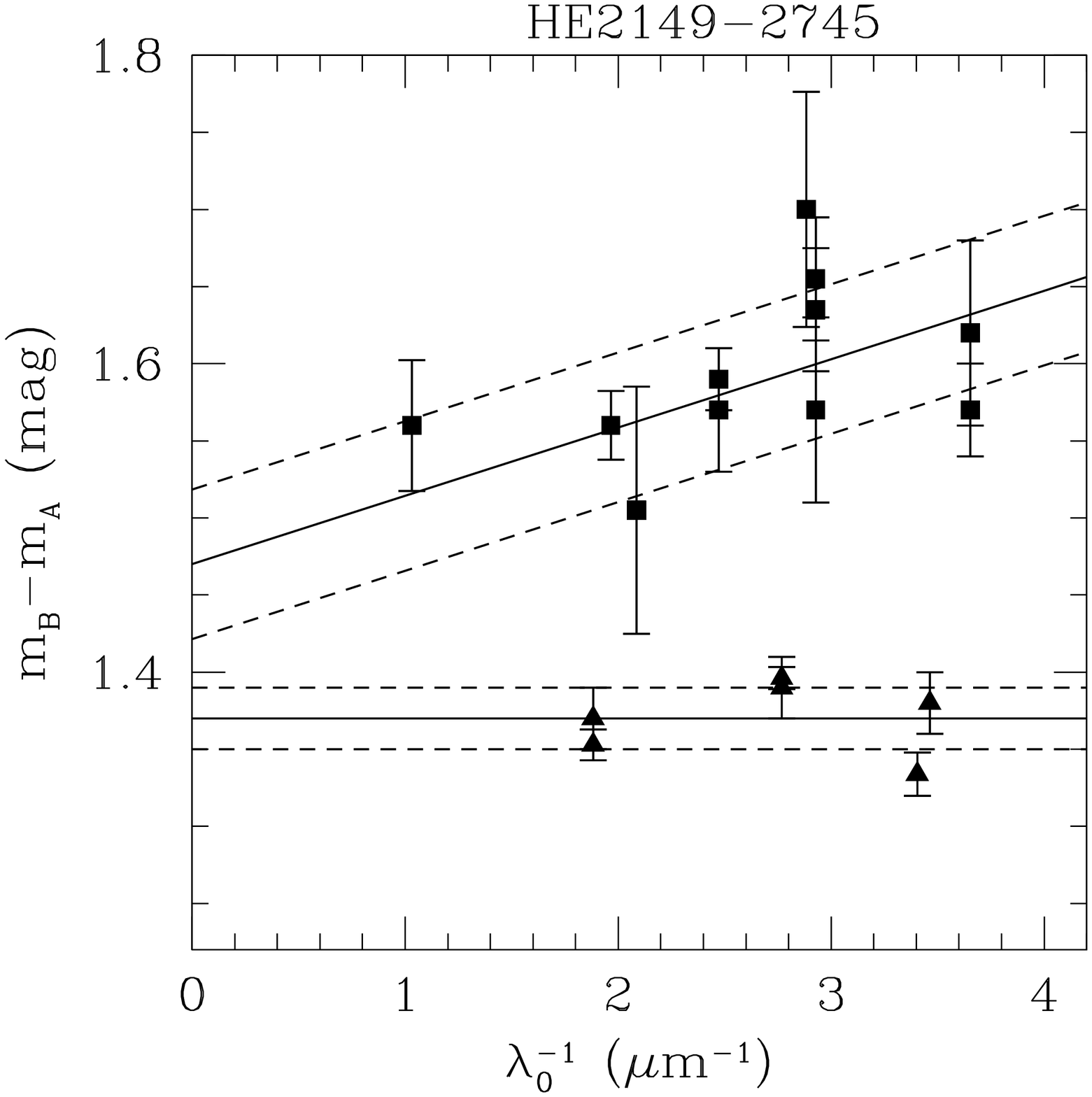}
\caption{Models fitted to the data shown in Figure \ref{diff_he2149a}.  {\em
  Squares} and {\em triangles} represent continuum and line core data
respectively.  {\em Black lines} represent the function fitted to the
continua and the average of the emission lines.  {\em Dashed lines}
are the standard deviations for the continuum fits and the standard
error of the mean for the emission lines.
\label{diff_he2149b}}
\end{center}
\end{figure*}

In Figure \ref{diff_he2149a} (Table \ref{mag_he2149}) we present the emission line and
continuum ratios for our data and for other data in the
literature. The emission line ratios show no dependence with
wavelength. The continuum flux ratios, however show chromaticity,
likely induced by microlensing (Figure \ref{diff_he2149b}).

\begin{deluxetable}{rc}
\tablecolumns{2}
\tablewidth{0pt} 
\tablecaption{HE2149-2745 chromatic microlensing \label{map_he2149}}
\tablehead{ 
  \colhead{$\lambda_c$ (\AA)} &
  \colhead{$\Delta m_C - \Delta m_L$\tablenotemark{a} (mag)} 
}
\startdata                  t
 4380    & $0.26\pm0.06$ \\ 
 8140    & $0.19\pm0.02$ \\ 
38000    & $0.15\pm0.01$ \\ 
\enddata
\tablenotetext{a}{Difference between the magnitude difference  $(m_B-m_A)_C-(m_B-m_A)_L$ in the broad-band data from CASTLES and $L$ band data from \cite{fadely11} and in the emission line cores from VLT data, re-analysis of \cite{sluse12} deconvolved spectra, and $Ks$ band data from \cite{fadely11} (see text).}
\end{deluxetable}

\begin{figure*}
\begin{center}
\plotone{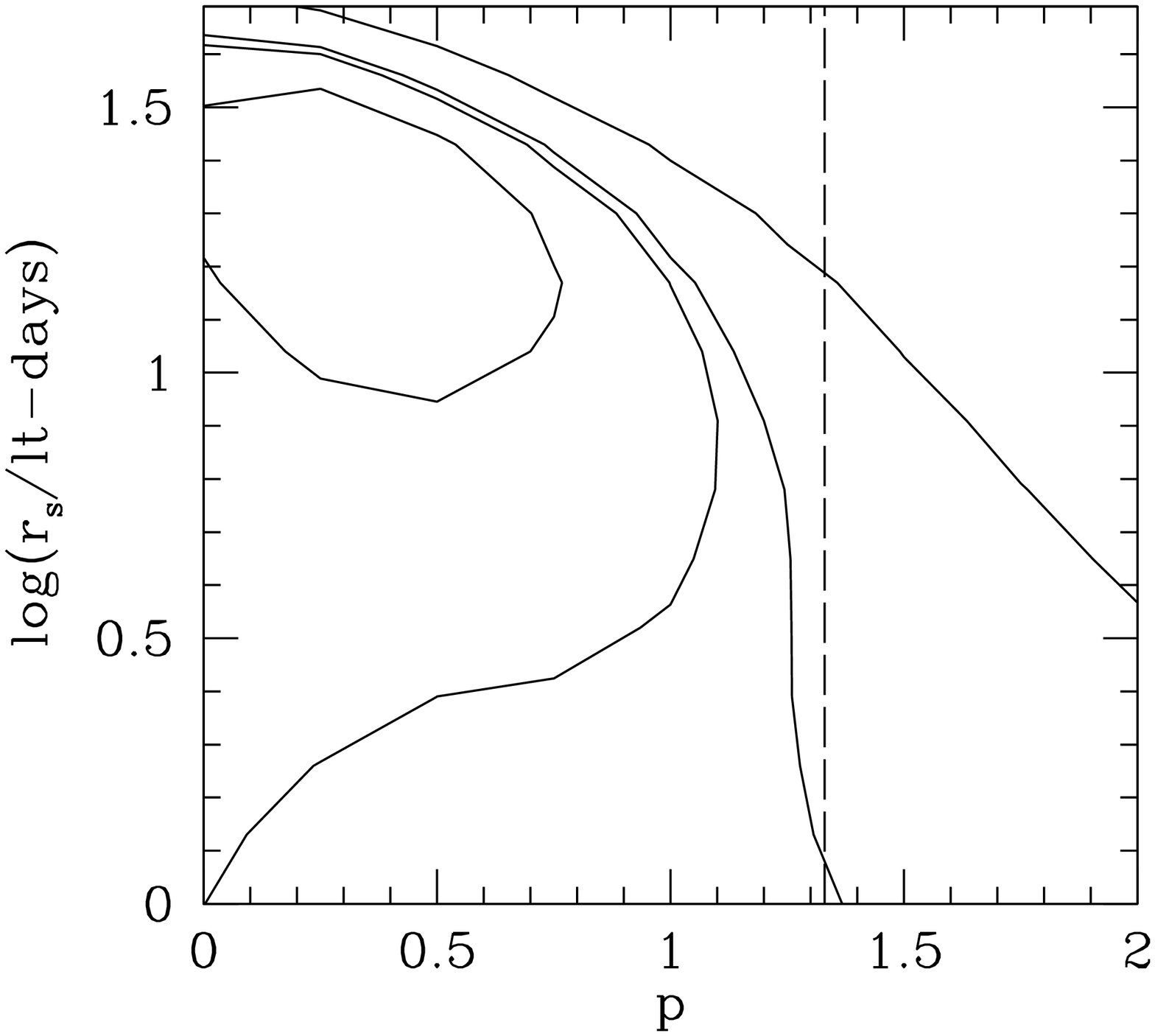}
\caption{Two-dimensional pdfs obtained using the measured chromatic 
microlensing for HE2149-2745 (Table \ref{map_he2149}) for 
 logarithmic grids in  $r_s$.
Contours correspond to $0.5 \sigma$, $1 \sigma$, $1.5 \sigma$, and 
$2 \sigma$ respectively.  The {\em dashed line} corresponds to the 
value predicted by  the thin disk model ($p=4/3$).
\label{he2149size}}
\end{center}
\end{figure*}

Using the continuum and emission line ratios, (see \S 3) we estimate
the size, $r_s$, and the logarithmic slope, $p$, of the size
dependence with wavelength. In this case we have used convergence $\kappa_A=0.31$,
 $\kappa_B=1.25$, and shear $\gamma_A=0.32$, $\gamma_B=1.25$ \citep{sluse12} to
compute the magnification maps. In Figure \ref{he2149size} we present
the 2D PDF of $r_s$ and $p$. The resulting estimates are
$r_s=8^{+11}_{-5}$ light-days (log prior) at $\lambda_0=1310$~\AA
\ and $p=0.4\pm0.3$ (linear prior).

\section{Conclusions}

We have used spectroscopy of three lensed quasars, HE0435-1223,
WFI2033-4723, and HE2149-2745, to study their inner structure (BLR and
continuum source) as derived from microlensing magnification. The
results are:

\begin{itemize}
\item
We have detected microlensing in the emission line profiles of two
of the lensed systems, HE0435-1223, WFI2033-4723. In the case of
HE0435-1223, we have found an enhancement of the red wing of image $D$
with respect to image $B$ in \ion{C}{4} and \ion{C}{3]} that, adopting the same
  criteria for line profile comparison, can be reconciled with
  previous results from infrared spectroscopy obtained by
  \cite{braibant14}. Using the measured microlensing magnification we
  estimate a size of $r_s=10^{+15}_{-7} \sqrt{M/M_{\odot}}$~light-days
  for the \ion{C}{4} emitting region affected by microlensing. In the case of
  WFI2033-4723, we have detected microlensing in two bumps situated on
  the blue and red wings of \ion{C}{3]}, confirming and extending previous
    microlensing evidence found in the blue wing \citep{sluse12}.  In
    principle, the blue bump might be associated to an \ion{Al}{3} emission
    line usually present in the red wings of the \ion{C}{3]} emission line.
      Alternatively, we could interpret the two bumps as evidence of
      the double-peaked profile typical of disk kinematics, assuming
      that part of the \ion{C}{3]} emission arises from the accretion
        disk. Using the measured microlensing magnification we
        estimate a size of $r_s=11^{+28}_{-7} \sqrt{M/M_{\odot}}$
        light-days for the region of the disk emitting the microlensed
        bumps. Combining this size with the velocity inferred from the
        wavelength separation between the bumps we obtain an estimate
        of $1.2^{+3.1}_{-0.8}\times10^8 M_\odot$ for the mass of the
        central supermassive black hole.
\item
The ratios of the line emission cores show no evidence of
chromaticity. This excludes both significant effects of microlensing
on the regions generating the cores of the emission lines, and
appreciable extinction.

\item
Using the continuum and (core) emission line ratios, we estimate
sizes of $13^{+5}_{-4}\sqrt{M/M_{\odot}}$,
$10^{+3}_{-2}\sqrt{M/M_{\odot}}$, $8^{+11}_{-5}
\sqrt{M/M_{\odot}}$~light-days, and slopes $1.2\pm0.6$, $0.8\pm0.2$,
and $0.4\pm0.3$ for HE0435-1223, WFI2033-4723, and HE2149-2745
respectively. In the case of HE0435-1223 and WFI2033-4723, the good
agreement between the sizes of the continuum and microlensed regions
in the emission line wings also support the hypothesis that the latter
arise from the accretion disk.

\item
The measured continuum microlensing amplitude (in the three
systems) and chromaticity (in WFI2033-4723 and HE2149-2745) are below
the predictions of the thin disk model. This results in larger disk
sizes and flatter temperature gradients than expected.

\end{itemize}

\acknowledgments

We thank the anonymous referee for useful suggestions.
V.M. gratefully acknowledges support from FONDECYT through grant
1120741 and Centro de Astrof\'{\i}sica de Valpara\'{\i}so.  E.M,
J.A.M. acknowledges support from MINECO and Junta de Andaluc\'{\i}a
through grants: AYA2011-24728,
AYA2013-47744-C3-1, AYA2013-47744-C3-3-P, and FQM-108.  JJV is supported by the project AYA2014-53506-P
financed by the Spanish Ministerio de Econom\'{\i}a y Competividad and
by the Fondo Europeo de Desarrollo Regional (FEDER), and by project
FQM-108 financed by Junta de Andaluc\'{\i}a.  JAM is also supported by the Generalitat Valenciana with the grant PROMETEO/2014/60. K.R. is supported by PhD fellowship FIB-UV 2015/2016. 
This research has made
use of NASA's Astrophysics Data System, and the NASA/IPAC Extragalactic Database (NED) which is operated by the Jet Propulsion Laboratory, California Institute of Technology, under contract with the National Aeronautics and Space Administration.

\end{document}